\documentclass[a4paper,fleqn,usenatbib]{mnras}

\usepackage[T1]{fontenc}
\usepackage{ae,aecompl}

\usepackage{graphicx}	
\usepackage{amsmath}	
\usepackage{amssymb}	
\usepackage[dvipsnames]{xcolor}

\usepackage{newtxtext,newtxmath}

\title[Transient behavior of SU UMa DNe]{Transient behavior of three SU UMa-type dwarf novae; AR Pic, QW Ser and V521 Peg}

\author[H. Szegedi et al.]{
H\'el\`ene Szegedi,$^{1}$\thanks{E-mail: SzegediH@ufs.ac.za}
Philip A. Charles,$^{2}$
Pieter J. Meintjes,$^{1}$
and Alida Odendaal$^{1}$
\\
$^{1}$Department of Physics, University of the Free State, 205 Nelson Mandela Drive, Bloemfontein, 9300, RSA\\
$^{2}$Department of Physics and Astronomy, University of Southampton, Southampton, Hampshire, SO17 1BJ, United Kingdom\\
}

\date{Accepted XXX. Received YYY; in original form ZZZ}

\pubyear{2020}

\begin{document}
\label{firstpage}
\pagerange{\pageref{firstpage}--\pageref{lastpage}}
\maketitle

\begin{abstract}
Changes in the supercycle lengths of some SU UMa-type dwarf novae have been detected by other studies, and indicate that the mass transfer rates noticeably decrease over time. We investigated the supercycle lengths of three SU UMa-type dwarf novae: AR Pic, QW Ser and V521 Peg, to determine if they have detectable changes in their supercycles. We present the results of optical spectroscopic and photometric observations of these sources. Our observations were conducted in 2016 and 2017 at the Boyden Observatory and the Sutherland station of the South African Astronomical Observatory. The quiescent results indicated that all three sources are typical SU UMa-type dwarf novae. We also present results of AR Pic and QW Ser in outburst and of V521 Peg during a precursor outburst and superoutburst. Light curves were supplemented by the Catalina Real-Time Transient Survey, the ASAS-3 and ASAS-SN archives, and the AAVSO International database in order to investigate the long-term behavior of these sources. Our results combined with catalogued properties for all short-period dwarf novae, shows a possible relationship between the supercycle time in SU UMa systems and their orbital periods, which is interpreted as the decline in the mass transfer rate as systems evolve toward and away from the `period minimum'. At the shortest orbital periods, SU UMa systems are almost indistinguishable from WZ Sge systems. However, we propose that the scaleheight between the secondary's photosphere and L1 may be a factor that distinguish the SU UMa subclasses.
\end{abstract}




\section{Introduction}

Cataclysmic variables (CVs) are close binary systems in which a Roche-lobe filling low-mass star transfers matter through the inner Lagrangian point (L1) to a white dwarf (WD). The magnetic field strength of the WD dictates the method by which the transferred material accretes onto the WD surface, which is either via an accretion disc (non-magnetic systems) or via the WD's magnetic field lines (magnetic systems). The WD primary has a mass range $0.8 ~{\rm M_{\sun}} \leq M_1 \leq 1.2~ \rm M_{\sun}$ (e.g. \citealp{Knigge2011a, Zorotovic2011}) and the secondary is a typical K- or M type star with a mass $M_2 \leq 0.5 \rm ~M_{\sun}$ (\citealp{Giovannelli2015, Smak1984a}). They have short orbital periods mostly in the 1--12 hr range, and hence are compact, with a binary separation $\sim \rm R_{\sun}$ (see \citealp{Warner1995} for an in-depth review).

Dwarf novae (DNe) are a non-magnetic CV sub-class and are the most common CV type, being characterized by optical outbursts ($\sim$2--8 mags) that recur on time-scales of days to years. SU UMa-type DNe are a further subclass that show superoutbursts in addition to the normal DNe outbursts, and uniquely display superhumps, i.e. prominent periodic humps in the light curves, during superoutbursts \citep[e.g][]{Vogt1974, Warner1975b}. 

Normal outbursts last for $\sim 2 - 3$ days and are caused by a build-up of material in the accretion disc that triggers a thermal-viscous instability \citep[e.g][]{Lasota01}. The disc switches from a cold, low-viscosity state (quiescence) to a bright, hot, highly viscous state (outburst) \citep[e.g.][]{Smak1984a}, in which the accretion rate onto the WD is greatly enhanced \citep{Osaki1974}. The DN returns to quiescence when a cooling wave, originating in the outer disc, propagates inwards and ends the outburst \citep[e.g.][p.~71, and references therein]{Hellier2001}.

Superoutbursts are less frequent than normal outbursts but last much longer ($\sim 10 - 14$ days). \cite{Whitehurst1988} discovered that these arise from a tidal instability in the disc of systems with a mass ratio $q = M_2/M_1 < 0.3$, which are almost entirely of short orbital periods ($\leq 2$hrs). In the context of the Thermal Tidal Instability Model (TTIM), the disc becomes tidally unstable when its outer radius exceeds a critical value of $R_{\rm cr} \simeq 0.46a$, and when a parametric resonance of 3:1 between the particle and binary orbit is reached \citep{Osaki1989, Hirose1990}. It therefore becomes unstable against non-axisymmetric perturbations and a slowly precessing, eccentric disc develops.

The low $q$ values of SU UMa-systems also makes them of interest as the WD analogue to the class of low-mass X-ray binary (LMXB) systems that contain accreting black-holes (BH), as they too have very low $q$ values ($\sim 0.1$), see e.g. \citet{Casares17}.  The BH LMXBs share other properties with SU UMa-systems, as they are also transients (mostly with recurrence times of $\sim$ years) and have even been seen to display superhumps at times during outburst \citep{DOD96}.

The exact trigger of SU UMa superoutbursts is still debated, but proposed models and observations show that normal outbursts are precursors of superoutbursts (see \citealp{Schreiber2004} for an in-depth review). During a normal outburst the disc expands as viscosity within the disc results in the outward transfer of angular momentum. When the disc expands far enough to interact with the secondary, the secondary's tidal torque is enhanced as angular momentum is transferred from the disc's outer edge to the secondary. Therefore, during every orbital cycle, the tidal influence of the secondary removes angular momentum from the disc's outer edge, causing it to contract radially. The surface density at the outer edge is hence pushed above a threshold, where no low-viscosity cool state exists, and a superoutburst is triggered as the disc is kept in a hot state (see \citealp{Osaki1989}, for a full analytical treatment and summary). A precursor (normal) outburst is typically seen as a shoulder on the rise to superoutburst \citep{Cannizzo2010} with a dip $\sim 0.25 - 3$ mag before the onset of superoutburst \citep[][p.~188, and references therein]{Warner1995}, and have been seen in many SU UMa DNe, e.g. VW Hyi \citep{Bateson1977}, T Leo \citep{Howell1999} and V1504 Cyg \citep{Osaki2013}.   

A superoutburst has a prolonged maximum state that is observable as a sloping plateau in light curves, and the decline to quiescence is slower than for normal outbursts. When the non-axisymmetric nature of the contracting disc decays and the tidal torque is reduced, the superoutburst ends when a cooling wave propagates inwards, resulting in the disc transitioning back to a low-viscosity state.

Superoutbursts are also characterized by superhumps in the light curves with amplitudes of $\sim 0.1 -0.3$ mag and periods $\sim 1 - 5$\% longer than $P_{\rm orb}$ \citep{Patterson2005}. These humps represent the varying tidal energy released by the deformation of the disc through an orbital cycle, but are typically delayed for $\sim 2$ days from the onset of the superoutburst (e.g. \citealp{Nogami2004a}, and references therein). The superhumps reach a maximum amplitude at outburst maximum and decay during the plateau phase. 

There are three distinct types of SU\,UMa DNe based on their supercycle lengths ($P_{\rm sc}$), i.e. the interval between two successive superoutbursts.  Apart from the `standard' SU\,UMa systems (with $P_{\rm sc} \sim 100 - 1000$ days), the two other types are at the extremes: ER\,UMa systems, which erupt frequently ($P_{\rm sc} \sim 20 - 50$ days), in contrast to WZ\,Sge systems, which have very rare superoutbursts ($P_{\rm sc} > 1000$ days), and also undergo few or no normal outbursts \citep[e.g.][p.~87, and references therein]{Hellier2001}. The mass transfer rate is the key factor that determines $P_{\rm sc}$, with typical values for SU\,UMa systems $\sim \dot{M_2} \sim 10^{-10} \rm ~M_{\sun}~yr^{-1}$. This increases to  $\sim \dot{M_2} \sim 6\times10^{-10} \rm ~M_{\sun}~yr^{-1}$ in ER\,UMa systems, resulting in more frequent superoutbursts, whereas it is much lower ($\dot{M_2} < 10^{-10} \rm ~M_{\sun}~yr^{-1}$) in WZ\,Sge types, hence producing much rarer superoutbursts.

\begin{figure*}
 \centering
 \includegraphics[width=0.8\textwidth]{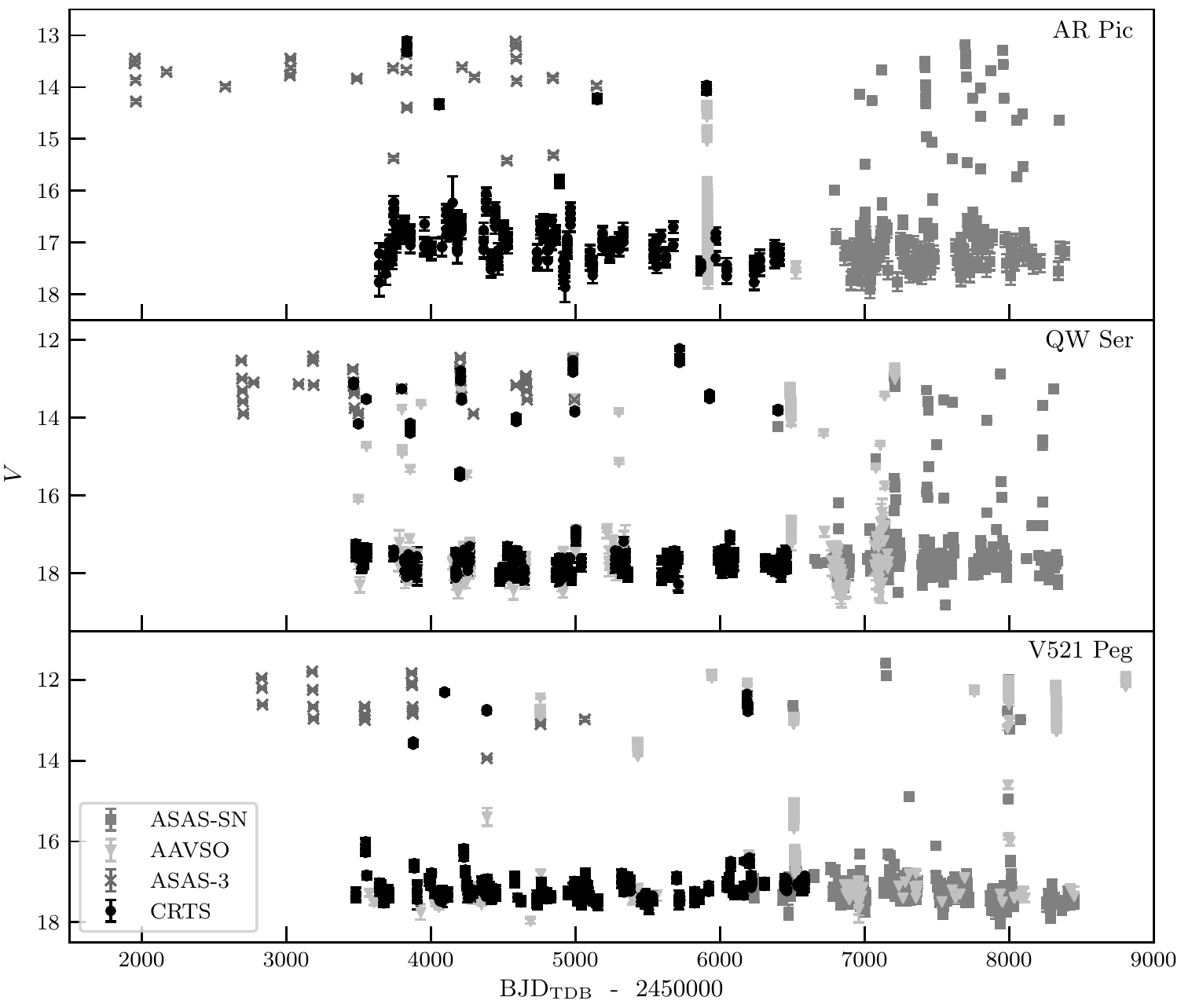}
 \caption{Composite light curves of AR Pic (\textit{top panel}), QW Ser (\textit{middle panel}) and V521 Peg (\textit{bottom panel}), compiled from the CRTS, ASAS-3, ASAS-SN and AAVSO archival data. The uncertainty in each data point is shown.}
 \label{longterm_lc}
\end{figure*}

These different $\dot{M_2}$ values are very likely a function of these systems' binary evolution, and \cite{Hypka2013} have shown that individual objects display increases in their $P_{\rm sc}$ values when observed over $\sim 20$ years. In an effort to expand on the list of SU\,UMa DNe showing such changes in $P_{\rm sc}$, we selected three systems for further study: AR Pic, QW Ser and V521 Peg. These have known $P_{\rm sc}$ values, are observable from our South African facilities, and indicate outburst activity in long-term light curves (as shown in Fig.~\ref{longterm_lc}'s $\sim 17$ year monitoring based on the CRTS \citep[Catalina Real-Time Transient Survey;][]{Drake2009, Drake2013, Drake2014}, ASAS-3 \citep[All Sky Automated Survey;][]{Pojmanski2002}, ASAS-SN \citep[All Sky Automated Survey for SuperNovae;][]{Shappee2014,Jayasinghe2019} and AAVSO \citep[American Association of Variable Star Observers;][]{Kafka2019} archives. 

We studied their normal and superoutburst recurrence timescales, and discuss these within the framework of current evolutionary models. The observational cadence of these surveys was low, and so we also performed our own observations (both photometry and spectroscopy), investigating our targets during normal outbursts (AR Pic and QW Ser) and a superoutburst (V521 Peg) to determine if their outburst properties deviated from previously published outbursts, and to search for changes in $P_{\rm sc}$.  V521 Peg is of particular interest since to date no normal outbursts have been recorded \citep{Rodriguez2005} and it undergoes infrequent superoutbursts \citep{Aung2006}. This behaviour resembles that of WZ Sge systems which prompted us to investigate V521 Peg's classification.

The known properties of AR Pic, QW Ser and V521 Peg are summarised in Section 2 and the observations, data reductions and supplementary data used are in Section 3. The results of the observations are presented in Section 4.

\section{Known properties}

The key properties of these three SU UMa-type DNe are summarised in Table~\ref{properties}, including, in particular, the latest Gaia DR2 distances. Given their relative proximity (a few hundred pc), we used the simple parallax equation ($d = \frac{1}{p}$) and associated errors ($\sigma(d) = \frac{\sigma(p)}{p^2}$) in computing the distances, rather than the more detailed analysis of \citet{Luri2018}.

\begin{table*}
\begin{center}
\small
\caption{Key properties of three SU UMa DNe.}
\label{properties}
\begin{tabular}{*{17}{@{~~}l@{~~~~~~~~~}c@{~~~~~~~~~}c@{~~~~~~~}c@{~~~~~~~}c@{~~~~~~~~~}cc@{~~~~~~~~~~}r@{~}c@{~}l@{~~~~~~~~}r@{~}c@{~}l@{~~~~~~~~}r@{~}c@{~}lc}}
\hline
Star & ~$V_{\rm q}$\textsuperscript{*} & ~$V_{\rm o}$\textsuperscript{*} & ~~~$P_{\rm orb}$\textsuperscript{**}  & ~~$P_{\rm sh}$\textsuperscript{**} & $P_{\rm c}$ & $P_{\rm sc}$ & \multicolumn{3}{l}{~~~$d$\textsuperscript{$\dagger$}} & \multicolumn{3}{c}{$K_1$} & \multicolumn{3}{c}{$\gamma$} & $i$ \\ 
 & & & (hr) & (hr) & (days) & (days) & \multicolumn{3}{l}{~~(pc)} & \multicolumn{3}{c}{($\rm km~s^{-1}$)} & \multicolumn{3}{c}{($\rm km~s^{-1}$)} & (\textsuperscript{$\circ$}) \\ 
\hline
AR Pic & 16.9 & 13.2 & 1.93 & 1.99 &  & 750 -- 800 & 319 & $\pm$ & 8 &  &  &  &  &  &  & \\
QW Ser & 17.9 & 12.6 & 1.79 & 1.85 & 50 & 220 -- 270 & 377 & $\pm$ & 37 & 85 & $\pm$ & 8 & 10 & $\pm$ & 6 & 60 \\
V521 Peg & 17.3 & 12.3 & 1.44 & 1.47 &  & 300 & 191 & $\pm$ & 3 & 50 & $\pm$ & 2 &  &  &  & 55 \\
\hline
\multicolumn{17}{l}{\footnotesize \textit{Note}: $V_{\rm q}$ = quiescent $V$, $V_{\rm o}$ = outburst $V$, $P_{\rm orb}$ = orbital period, $P_{\rm sh}$ = superhump period, $P_{\rm c}$ = duty cycle, $P_{\rm sc}$ = supercycle,}\\
\multicolumn{17}{l}{\footnotesize $d$ = distance, $K_1$ = radial velocity semi-amplitude of primary, $\gamma$ = systemic velocity, $i$ = inclination.}\\
\multicolumn{17}{l}{\textsuperscript{*}\footnotesize{~CRTS quiescent (q) and outburst (o) $V$ \citep{Drake2014}.}}\\
\multicolumn{17}{l}{\textsuperscript{**}\footnotesize{Orbital ($P_{\rm orb}$) and superhump ($P_{\rm sh}$) periods from the Outburst Catalogue of CVs \citep{Coppejans2016}.}}\\ 
\multicolumn{17}{l}{\textsuperscript{$\dagger$}\footnotesize{~From ESA Gaia DR2 parallaxes \citep{Luri2018}.}}\\
\multicolumn{17}{l}{All other properties referenced in text.}\\
\end{tabular}
\end{center}
\end{table*}

\subsection{AR Pic} 
\label{arpic_info}
AR Pic (CTCV J0549-4921) is a known SU UMa-system (e.g. \citealp{Maza1989, Tappert2004, Imada2008}) with an average $V \sim 16.9$ in quiescence, during which it displays typical DN features, e.g. moderately strong H and He{\sc~i}, and fainter He{\sc~ii} and Fe{\sc~ii} emission lines \citep{Tappert2004}. Quiescent light curves exhibit two unequal humps which are attributed to an observable hot spot typically seen in low mass transfer rate DNe (see \citealp{Tappert2004}, and references therein). They also show no eclipses, indicating that AR Pic has a low to intermediate inclination \citep{Imada2008}.
\cite{Tappert2004} determined an orbital ephemeris, but its phasing is now indeterminate.

\cite{Kato2009} determined that AR Pic generally has short outbursts that only last a day, with an average outburst amplitude of $\sim 4$ mag \citep{Drake2009}.  It showed no periodic features during outburst, but moderate flickering \citep{Tappert2004}.

\cite{Imada2008} were the first to classify AR Pic as an SU UMa-system, after observing a superoutburst in 2006 that was detected during the rising phase, reaching a maximum of $V \sim 13$. It declined during the plateau phase at a rate of $0.12\pm0.01 \rm ~mag~day^{-1}$, and showed superhumps with a period of $P_{\rm sh} = 2.024\pm0.004$ hr, yielding a period excess of 
\begin{equation}
   \varepsilon = \frac{P_{\rm sh} - P_{\rm orb}}{P_{\rm orb}} \simeq 0.05.
\end{equation}
Using ASAS-3 light curves, \cite{Kato2009} determined that the superoutbursts' duration vary between 6 and 10 days and that the supercycle length is approximately $750 - 800$ days. \cite{Kato2013} deduced a mean $P_{\rm sh} = 1.996\pm0.004$ hr from data obtained during a superoutburst in 2011, i.e. a period excess of $\varepsilon \simeq 0.037$

\subsection{QW Ser}

QW Ser (TmzV46) was a known DN (e.g. \citealp{Takamizawa1998, Schmeer1999, Kato1999}) and later confirmed as SU UMa-type (e.g. \citealp{Patterson2003, Olech2003, Nogami2004a}).  It is quite faint during quiescence ($V \sim 18$) and exhibits a double hump structure in the light curves, attributed to orbital modulation \citep{Olech2003}.  Its quiescent spectrum exhibits double-peaked emission lines from the accretion disc \citep{Patterson2003}. The H$\alpha$ line, with a peak separation of $ \sim 1100  \rm~km~s^{-1}$, was used to determine a systemic velocity of $\gamma = 10\pm6 \rm ~km~s^{-1}$, and a radial velocity semi-amplitude of $K_1 = 85\pm8 \rm ~km~s^{-1}$ for the WD. \cite{Patterson2011} determined an inclination angle of $i \simeq 60^{\circ}$.

Previous outbursts in ASAS-3 \citep{Takamizawa1998} show outburst durations $\sim 2 - 3$ days (see e.g. \citealp{Nogami2004a}), however, \cite{Kato1999} reported one lasting $\sim 11 - 16$ days. \cite{Nogami2004a} determined a maximum $M_{\rm V}$ of $5.2\pm0.2$ and normal outburst duty cycle of $\sim 50$ days. During superoutbursts, QW Ser typically had an outburst amplitude of $\sim 5$ mag and lasted $\sim 15$ days (e.g \citealp{Patterson2003, Olech2003, Nogami2004a}). It slowly declined during the plateau phase at a rate of $0.086-0.11 \rm ~mag~day^{-1}$ and rapidly thereafter at $1.2 \rm ~mag~day^{-1}$. Superhumps developed within 2 days of the superoutburst onset and reached a maximum amplitude of 0.35 mag within 4 days of onset \citep{Nogami2004a}. Since 1999, the supercycle length has been $\sim 220 - 270$ days.

\subsection{V521 Peg}

V521 Peg (HS 2219+1824) was first classified as a CV candidate in the Hamburg Quasar Survey \citep{Gansicke2002}, and later identified as a SU UMa-system \citep[hereafter RG05]{Rodriguez2005}. It has a quiescent $V \sim 17.3$ and displays a double hump structure of uneven amplitude that was associated with an observable hot spot \citep[RG05,][]{Aung2006}. 

In quiescence, it displayed double-peaked H and He{\sc~i} emission lines from a disc, indicating an average velocity of $\sim 660 \rm ~km~s^{-1}$ for the outer disc edge, but no absorption features of the donor star were detected (RG05). Broad Balmer absorption troughs, in which the H$\beta$ to H$\gamma$ emission lines were embedded, were attributed to the WD photospheric spectrum. RG05 also detected a peculiar, very narrow emission feature within the double-peaked profile which they attributed to chromospheric emission from the inner face of the secondary, possibly due to UV-irradiation from the hot WD. Velocities of $K_1 = 50\pm2 \rm~km~s^{-1}$ for the WD and $K_2 = 257\pm10 \rm~km~s^{-1}$ for the secondary star were determined from H$\alpha$ emission radial velocity curves and \cite{Patterson2011} estimated $i \simeq 55^{\circ}$. 

To date, no normal outbursts had been recorded for this source. V521 Peg is regarded as a ``low-activity'' system that undergoes infrequent superoutbursts \citep{Aung2006}, with supercycle estimates of 300 days \citep{Patterson2011} and 170 days \citep{Hypka2016}.  Superoutbursts had amplitudes of $\sim 5$ mag and lasted longer than $\sim 10$ days (RG05). Superhumps, with $P_{\rm sh} \simeq 1.484$ hr, were observed approximately 1 day after superoutburst maximum, and Kato et al. ($2014b$) \nocite{Kato2014b} noted that the superhump profile in the light curves of a 2013 superoutburst became double-humped during the rapid fading phase.

\section{Observations and data reductions}

Optical spectroscopy and photometry of our three targets was conducted between June 2016 and September 2017, of which 40\% were performed simultaneously.  Supplementary data was obtained from the CRTS, ASAS-3, ASAS-SN and AAVSO archives.   

\subsection{Spectroscopy}

Optical spectroscopy of these sources was conducted at the Sutherland station of the South African Astronomical Observatory (SAAO), using the 1.9-m telescope equipped with SpUpNIC (Spectrograph Upgrade - Newly Improved Cassegrain). For details of these observations and the source states, see Table~\ref{spec_obs}. Gratings G4 and G7, with resolutions of 1.8~\AA~and 6.6~\AA, respectively \citep{Crause2016a}, were utilized, together with a CuAr lamp for wavelength calibration with both gratings. The slit width was set to $1.5''$ for all observations and no star filters were used. 

\begin{table*}
\begin{center}
\footnotesize
\caption{Log of spectroscopic observations conducted with the SAAO 1.9-m telescope and SpUpNIC.} 
\label{spec_obs}
\begin{tabular}{llccccl}
\hline
Target & Date & HJD & Grating & $n_{\rm data} \times t_{\rm exp}$ & $\Delta \lambda$ & State\\ 
 & & (2450000+) & & (s) & (\AA) & \\
\hline
AR Pic & 2017-02-15 & 7800 & G7 & 2$\times$1800 & 3900 - 9600 & Quiescence \\
 & 2017-02-18, 19 & 7803-7804 & G7 & 2$\times$1800, 1$\times$1500 & 3500 - 9200 & Outburst - Fading \\
 & 2017-02-20, 21, 23 & 7805-7808 & G7 & 5$\times$1500, 1$\times$1800 & 3500 - 9200 & Quiescence \\
 & 2017-02-22, 26 & 7807-7811 & G4 & 1$\times$600, 6$\times$900, 13$\times$1000 & 3900 - 5200 & Quiescence \\
QW Ser & 2016-08-04 & 7605 & G7 & 2$\times$1800 & 2800 - 8500 & Quiescence\\ 
 & 2016-08-08, 09, 10 & 7609-7611 & G4 & 41$\times$600 & 3800 - 5100 & Outburst - Fading\\
 & 2016-08-15 & 7616 & G7 & 2$\times$1800 & 3450 - 9150 & Quiescence\\ 
 & 2017-02-23 & 7808 & G7 & 2$\times$1800 & 3650 - 9350 & Quiescence\\
 & 2017-08-26 & 7992 & G7 & 1$\times$1800 & 3600 - 9200 & Quiescence\\
V521 Peg & 2016-08-09 & 7610 & G4 & 4$\times$1300 & 3800 - 5100 & Quiescence \\
 & 2017-08-27 & 7993 & G7 & 3$\times$1200 & 3600 - 9200 & Precursor outburst \\ 
  & 2017-09-01, 03, 04 & 7998-8001 & G7 & 12$\times$400, 1$\times$1200 & 3600 - 9200 & Superoutburst \\
 & 2017-09-02, 03, 04 & 7999-8001 & G4 & 21$\times$600, 23$\times$400 & 3900 - 5200 & Superoutburst \\  
\hline
\multicolumn{7}{l}{\footnotesize{\textit{Note}: Slit width was $1.5''$ for all observations, giving spectral resolutions of 1.8~\AA ~(G4) ~and 6.6~\AA ~(G7).}}\\
\multicolumn{7}{l}{\footnotesize{~~~~~~~~~The number of data frames is $n_{\rm data}$ and the exposure time per frame is $t_{\rm exp}$.}}\\
\end{tabular}
\end{center}
\end{table*}

\begin{table*}
\begin{center}
\footnotesize
\caption{Log of photometric observations conducted at the Boyden Observatory and SAAO.} 
\label{phot_obs}
\begin{tabular}{lllcl@{~}l@{~}l@{~}l@{~}lcl}
\hline
Target & Date & Telescope & HJD & \multicolumn{5}{c}{Filter} & $t_{\rm exp} \times n_{\rm data}$ & State\\ 
 & & & (2450000+) & \multicolumn{5}{c}{} & (s) & \\
\hline
AR Pic & 2017-02-19 & 1.0-m/SAAO & 7804 & & & C & & & 5$\times$1441 & Outburst - Fading\\
  & 2017-02-22, 23, 26 & 1.0-m/SAAO & 7807-7811 & & & C & & & 5$\times$7323  & Quiescence\\
QW Ser & 2016-08-04 & 1.5-m/Boyden & 7605 & & & C & & & 25$\times$238 & Quiescence\\
 & 2016-08-08, 09, 10, 11, 12 & 1.5-m/Boyden & 7609-7613 & & & C & & & 25$\times$1501, 40$\times$255 & Outburst - Fading\\
 & 2016-08-10, 11, 12 & 0.4-m/Watcher & 7611-7613 & & $\rm g'$, & $\rm r'$, & $\rm i'$ & & 10$\times$416, 30$\times$185 & Fading\\
 & 2016-08-14 & 1.5-m/Boyden & 7615 & & & C & & & 40$\times$310 & Quiescence\\
 & 2016-08-14 & 0.4-m/Watcher & 7615 & & $\rm g'$, & $\rm r'$, & $\rm i'$ & & 30$\times$152 & Quiescence\\
 & 2017-03-04 & 1.0-m/SAAO & 7817 & & & C & & & 30$\times$306 & Quiescence\\ 
V521 Peg & 2016-08-08, 09 & 1.5-m/Boyden & 7608-7609 & & & C & & & 30$\times$221 & Quiescence\\
 & 2017-09-02, 03, 04, 05 & 1.5-m/Boyden & 7999-8002 & & & C & & & 10$\times$1955, 20$\times$1903 & Superoutburst\\
 & 2017-09-04, 05, 06, 07, 08, 09 & 0.4-m/Watcher & 8001-8006 & B, & V, & R, & I & & 30$\times$888 & Superoutburst - Fading\\ 
 & 2017-09-10, 11-14, 16-21, 23, 27 & 0.4-m/Watcher & 8007-8024 & B, & V, & R, & I, & C & 30$\times$1458 & Quiescence\cr 
\hline
\multicolumn{11}{l}{\footnotesize{\textit{Note}: The exposure time per frame is $t_{\rm exp}$, and $n_{\rm data}$ is the number of data frames.}}\\
\end{tabular}
\end{center}
\end{table*}

Standard routines of {\small IRAF} were used to perform bias and flat field correction, and the spectra were calibrated in wavelength using the CuAr arc frames.  Flux calibration was performed with respect to the spectrophotometric standard stars observed in the same night. The cosmic ray removal performed by the \textit{apall} task in {\small IRAF} was sufficient and the spectra of each night were stacked to obtain one average spectrum per date. Note that emission lines marked as H$\epsilon^*$ in the given spectra, are H$\epsilon$ lines blended with Ca{\sc~ii} lines, and the $\earth$ symbol marks the atmospheric A- and B-bands.

\subsection{Photometry}
\label{photometry}

Photometric observations were conducted at the Boyden Observatory, Bloemfontein, with the UFS-Boyden 1.52-m and 0.4-m Watcher robotic telescope\footnote{Operated by the Space Science Group at the University College Dublin \citep{Ferrero2010, Hanlon2013}}, and at the SAAO, Sutherland, with the 1.0-m plus SHOC (Sutherland High-speed Optical Camera).  The details of these observations are summarized in Table~\ref{phot_obs}. 

The UFS-Boyden 1.52-m telescope was equipped with a $770 \times 1152$ pixel Apogee U55 Back-illuminated CCD camera that provided a field-of-view (FOV) of $2.5' \times 3.7'$, and we used a C-filter (Clear, white light). After an outburst of QW Ser (August 2016) and a superoutburst of V521 Peg (September 2017) were detected, the Watcher robotic telescope was utilized to continue monitoring them until they returned to quiescence.  Watcher was equipped with a $1024 \times 1024$ pixel Andor iXon$+$ 888 CCD camera and a Meade $f/6.3$ Series 4000 focal reducer, providing a FOV of $10' \times 10'$ \citep{Murphy2014}. The QW Ser outburst was followed with $\rm g', i', r'$ filters and the V521 Peg superoutburst with Bessel BVRI and C filters. Only the C-filter and $\rm r'$-filter data were used for this study.

The SAAO 1.0-m telescope was equipped with SHOC, a high-speed instrument that can achieve frame rates of up to 20 frames per second during normal operations. SHOC\footnote{Further details on SHOC can be obtained from {https://www.saao.ac.za/science/facilities/instruments/shoc/}.} consists of an Andor iXon X3 888 Electron Multiplying (EM) $1024 \times 1024$ pixel back-illuminated CCD camera, providing a FOV of $2.85' \times 2.85'$, together with a global positioning system (GPS), and two filter wheels that include Bessel UBVRI, Sloan $\rm u'g'r'i'z'$ and Stromgren uvby filters. Our observations were performed in conventional mode i.e. utilizing a 1 MHz 16 bit conventional amplifier, and the internal computer clock was used to trigger the camera and provide time stamps. No filters were used and we only observed the sources in ``white light''. 

Data reductions, namely bias, dark current and flat-field corrections, were performed by using standard routines within the {\small IRAF} package \citep{Massey1997}. A Python-based pipeline\footnote{https://shoc.saao.ac.za/Pipeline/} developed by \cite{Kotze2013} was partially used to perform reductions on SHOC data. The {\small DAOPHOT} package in {\small IRAF} was used to apply aperture photometry. A weighted differential photometry procedure was applied, following the same method as discussed by \cite{Everett2001} and \cite{Burdanov2014}. Two comparison stars with the lowest mean variance in magnitude were utilized. Differential magnitudes were not transformed to a standard system for absolute photometry. For different observations to be comparable, a magnitude offset was determined for each dataset and added to the differential magnitudes. These offset values were calculated by using the average offset of the comparison stars' magnitudes from those in the NOMAD Catalog \citep{Zacharias2005}. The $V$-magnitude offset values were then applied to the differential magnitudes. For each exposure the mid-exposure time was converted to Barycentric Julian Date (BJD) in the Dynamical Time standard (TDB), using the `UTC2BJD' Time Utility of the Ohio State University\footnote{http://astroutils.astronomy.ohio-state.edu/time/utc2bjd.html} \citep{Eastman2010}. 
 
\subsection{CRTS, ASAS-3, ASAS-SN and AAVSO data}

Light curves from the CRTS, ASAS-3, ASAS-SN and AAVSO archives were used to increase the photometric baseline for our targets. The benefit of utilizing long-term light curves from various surveys is that outbursts possibly missed by one survey could have been detected by another. 

The CRTS Data Release 2\footnote{http://nesssi.cacr.caltech.edu/DataRelease/}, and 3\footnote{https://crts.iucaa.in/CRTS/}, and the ASAS-3\footnote{http://www.astrouw.edu.pl/asas/?page=aasc\&catsrc=asas3} websites provide magnitudes already transformed to Landolt $V$. The JDs of the downloaded CRTS data were converted to $\rm BJD_{TDB}$ using the same method described in Section~\ref{photometry}. We used the ASAS-3 data extracted with aperture 3 (`MAG\_2') as it had the lowest errors, whereas data with `C' (=`29.99, not measured') and `D' (=`worst data, probably useless') quality flags were removed. For each ASAS-3 data point the Heliocentric Julian Date (HJD) was converted to $\rm BJD_{TDB}$ by using the `HJD2BJD' Time Utility\footnote{http://astroutils.astronomy.ohio-state.edu/time/hjd2bjd.html} \citep{Eastman2010}.

The ASAS-SN $V$-band data\footnote{https://asas-sn.osu.edu/variables} were downloaded and the times converted to $\rm BJD_{TDB}$ in the same way. The AAVSO International Database is a collection of variable star observations made primarily by the amateur astronomical community, but also includes data from professional researchers and published literature. AAVSO technical staff prepare and maintain the database through strict quality-control analysis, and perform data error checks at every stage.  We downloaded our V-band (or `CV', meaning `Clear reduced to V sequence') data\footnote{https://www.aavso.org/data-download}, removing any that did not have a `V' (=`observation passed validation tests') or `Z' (=`pre-validation, data checked for typos and input errors') flag, and times were again converted to $\rm BJD_{TDB}$ as described above.

All the above data were used to plot a composite light curve for each source from which the duty cycles and supercycles could be determined (Fig.~\ref{longterm_lc}).

\section{Results}

\subsection{AR Pic}
\label{arpic_section}
AR Pic was observed by CRTS, ASAS-3, ASAS-SN and AAVSO in quiescence and during approximately 24 outbursts. The 2001--18 light curve (Fig.~\ref{longterm_lc}, top panel), indicates an average $V_{\rm q} \sim 17$ and outburst amplitude of $\sim 4$ mag, as seen earlier \citep{Drake2009}. By examining each outburst individually for duration, mean decline rate (outburst $\sim 0.6-0.8$ mag $\rm day^{-1}$; superoutburst $< 0.5$ mag $\rm day^{-1}$) and detectable plateau phase (decline rate $\sim 0.1$ mag $\rm day^{-1}$), we identified 8 normal and 7 superoutbursts. Due to the low cadence it was not possible to detect superhumps or the nature of all outbursts. From this we estimate a duty cycle of $\sim 60$ days and $P_{\rm sc} \sim 510$ days. Our $P_{\rm sc}$ is shorter than the $P_{\rm sc} \sim 750 - 800$ days determined by \cite{Kato2009}.  However, we do not ascribe this to a change in the system's mass transfer rate, but rather a better constraint of $P_{\rm sc}$, since our light curve spans a longer interval than the ASAS-3 light curve used by Kato et al.

\begin{figure}
 \centering
 \includegraphics[width=0.475\textwidth]{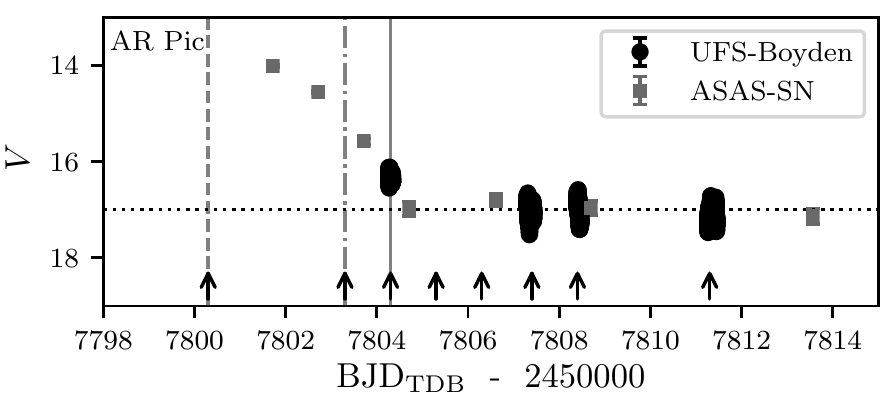}
 \caption{Overall light curve of AR Pic's Feb 2017 outburst, supplemented by ASAS-SN. Error bars are smaller than the symbols. Arrows indicate dates when spectra were obtained.}
 \label{arpic_outburst}
\end{figure}

During spectroscopic observations on 18 February 2017 (Fig.~\ref{arpic_spec}, top panel), the source was found to be $\sim 2$ mags brighter than 3 days earlier, and was in outburst. Fig.~\ref{arpic_outburst} shows the composite outburst light curve compiled from the UFS-Boyden 1.52-m telescope data and supplemented with ASAS-SN data. The black arrows indicate where spectra were obtained and show that the quiescent spectrum on 15 February (dashed line) was taken just before the rise to outburst. The spectrum from 18 February was obtained $\sim 3$ days into outburst (dot-dash line). AR Pic reached quiescence within 5 days, fading at a mean rate of 0.74 mag $\rm day^{-1}$. These observations confirm that AR Pic has normal outbursts that last longer than $\sim 1$ day, contrary to previous reports \citep{Kato2009}.

\begin{figure}
 \centering
 \includegraphics[width=0.47\textwidth]{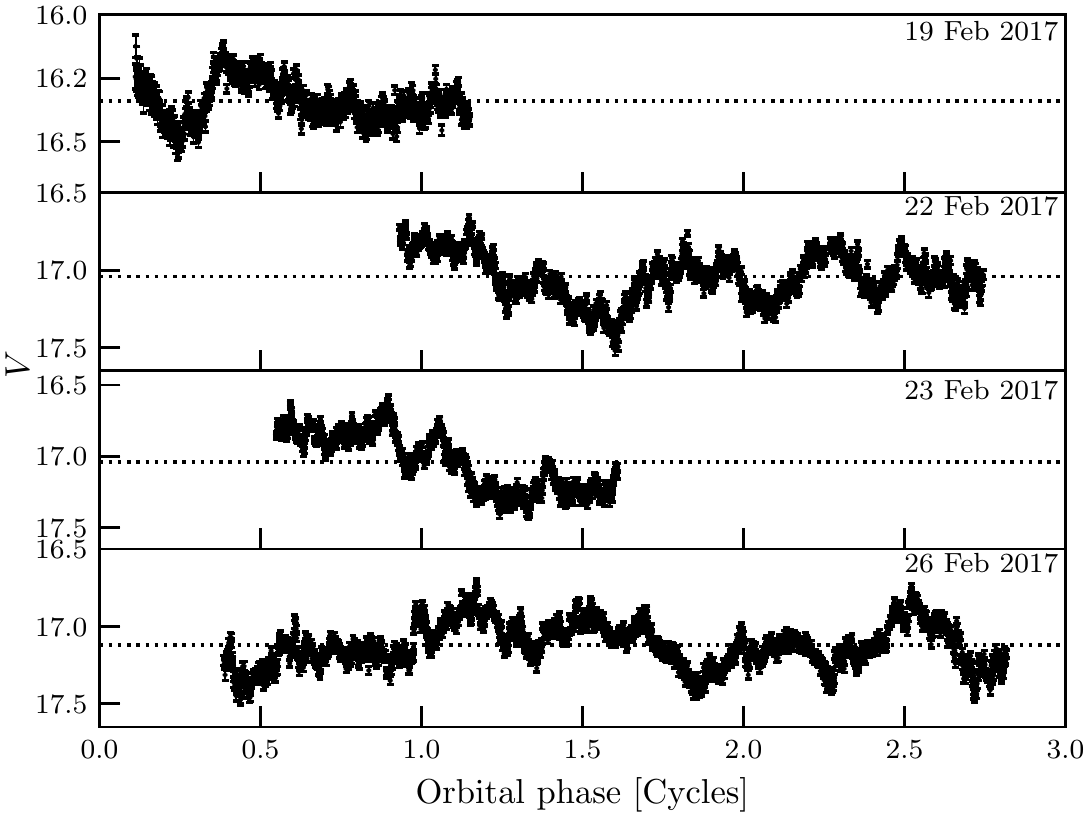}
 \caption{The light curves of AR Pic obtained during an outburst (19 Feb 2017) and during quiescence (22, 23, 26 Feb 2017). They show prominent flickering, but no orbital modulation is visible. Error bars are included but small relative to the data points.}
 \label{arpic_flick}
\end{figure}

To establish the nature of the outburst, simultaneous spectroscopic and photometric observations were conducted on 19 February 2017 (gray line in Fig.~\ref{arpic_outburst}). No superhumps were detected in the light curve (Fig.~\ref{arpic_flick}, top panel), but this is clarified by Fig.~\ref{arpic_outburst} which shows that the photometric data were obtained far past the peak of the outburst. There is also no plateau phase in Fig.~\ref{arpic_outburst}, therefore, the outburst was classified as a normal one. The emission lines were notably single-peaked during the outburst (Fig.~\ref{arpic_spec}, top panel), indicating that the outburst region dominated the emission. 

\begin{figure}
 \centering
 \includegraphics[width=0.45\textwidth]{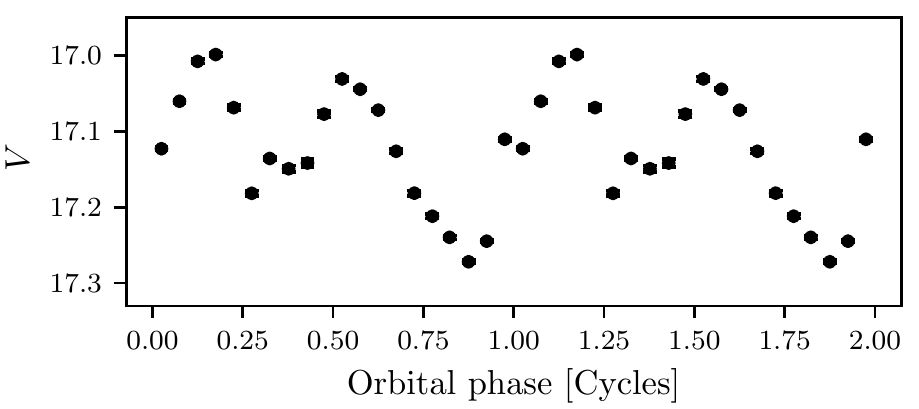}
 \caption{The quiescent light curve of AR Pic (26 Feb 2017), folded on the \citet{Tappert2004} ephemeris. Error bars are smaller than the symbols. Two cycles are shown for clarity.}
 \label{arpic_folded}
\end{figure}

\begin{figure*}
 \centering
 \includegraphics[width=0.8\textwidth]{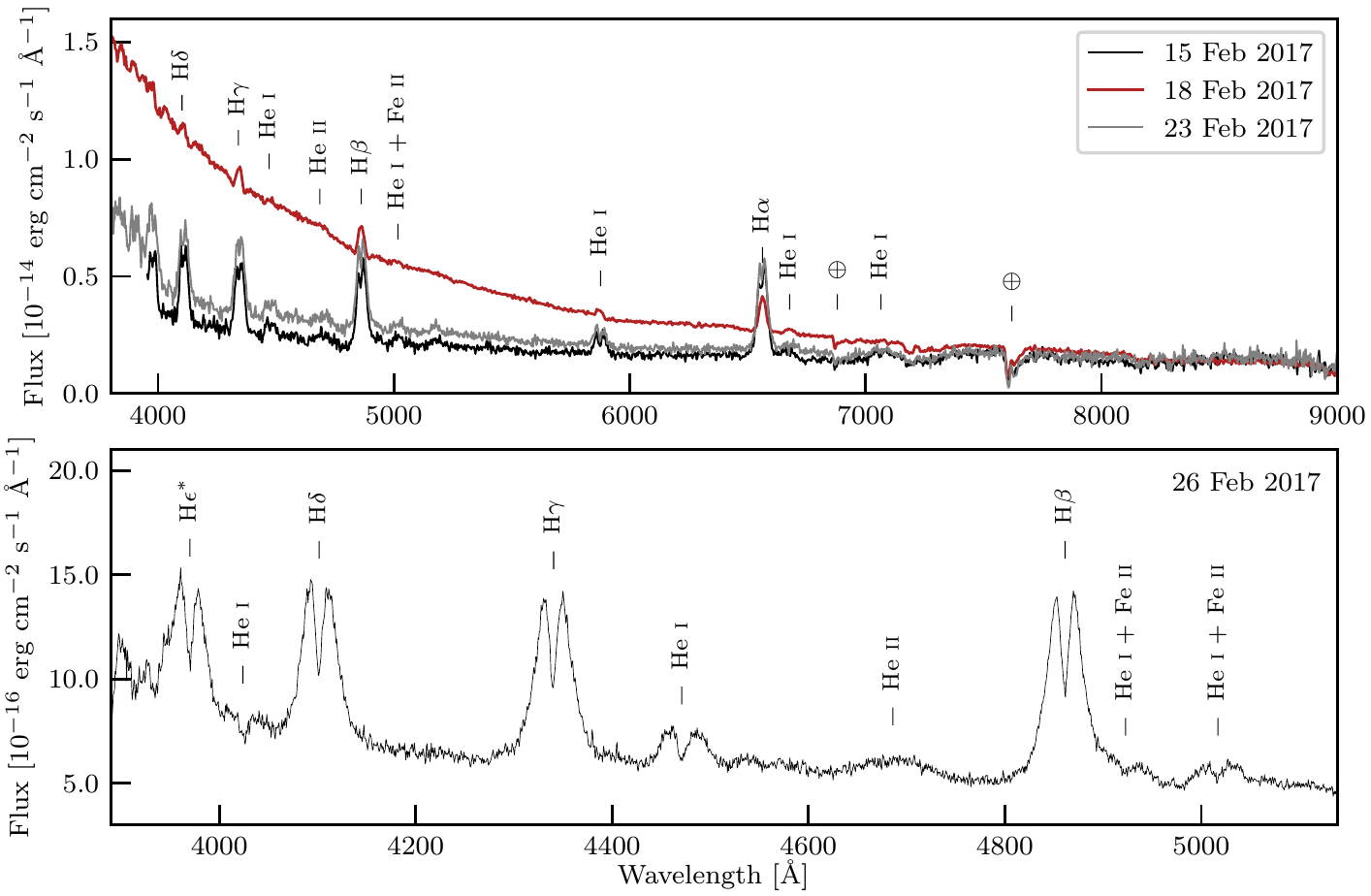}
 \caption{\textit{Top panel}: Spectra of AR Pic in quiescence (15 Feb 2017), outburst (18 Feb 2017) and after returning to quiescence (23 Feb 2017). The spectra of 15 and 23 Feb are multiplied by a factor of 10 to amplify features. \textit{Bottom panel}: A higher resolution quiescent spectrum clearly showing the double-peaked emission profiles.}
 \label{arpic_spec}
\end{figure*}

The known ephemeris of AR Pic (see \S\ref{arpic_info}; converted to $\rm BJD_{TDB}$) was used to calculate the orbital phases in Fig.~\ref{arpic_flick}. No eclipses were detected in these light curves, indicating an inclination of $i < 75^{\circ}$ (e.g. \citealp[p.~190]{Hellier2001}, and references therein). The strong flickering in the light curves, also observed by \cite{Tappert2004}, is further evidence of a very active hot spot and possible magnetic eddies that formed in the disc. The light curves do not appear to show an obvious correlation between  flickering and orbital phase. This was investigated by binning the photometric data of 26 February into 0.05 phase bins and folding on the ephemeris given by \cite{Tappert2004}. Two humps are visible in the folded light curve shown in Fig.~\ref{arpic_folded} and suggests that flickering is superimposed on the orbital modulation. However, the minimum in the light curve has shifted, as expected given the original ephemeris accuracy. The two humps were also detected by \cite{Tappert2004} and correlates with the result of \cite{Dai2018}, who modelled the contribution of hotpots to the optical emission of accretion-powered systems. 

In quiescence, AR Pic displayed a typical DN spectrum with H, He{\sc~i} and Fe{\sc~ii} emission (Fig.~\ref{arpic_spec}, 15, 23 and 26 Feb 2017), along with double-peaked profiles indicating the presence of an accretion disc. No WD or donor features were visible. 

\begin{figure}
 \centering
 \includegraphics[width=0.6\columnwidth]{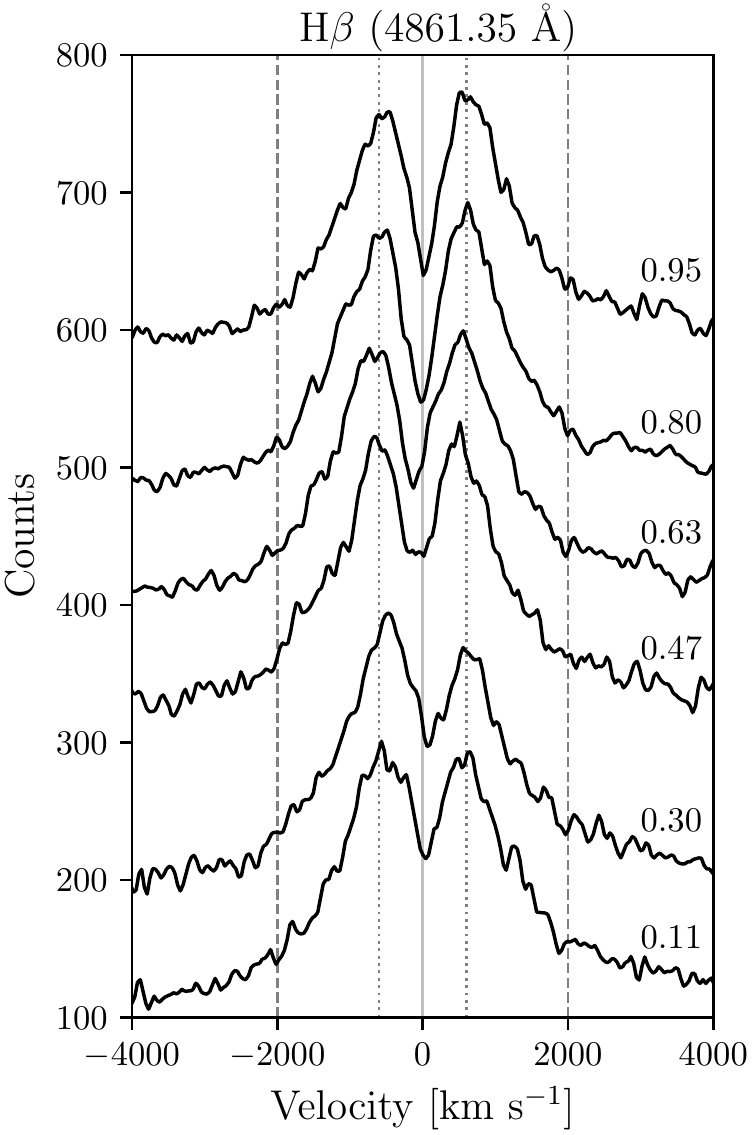}
 \caption{H$\beta$ profiles in binned spectra of AR Pic on 26 Feb 2017 show a changing asymmetry of the double peak, likely due to an accretion hot spot. The plots are offset by an arbitrary constant and are in phase order (labelled on the right). The dotted lines indicate the peak velocities, giving a peak separation of $\sim 1200 \rm ~km~s^{-1}$; the dashed lines indicate the line wings at $\sim 2000 \rm ~km~s^{-1}$.}
 \label{arpic_beta}
\end{figure}

Fig.~\ref{arpic_beta} illustrates the changing asymmetry of the double-peaked H$\beta$ profile in phase-combined spectra from 26 February 2017. The asymmetrical change is attributed to the hot spot rotating with the orbital period. The hot spot is clearly visible at phase 0.3 (blue peak higher) and at phase 0.8 (red peak higher). The peak separation is $\sim 1200 \rm ~km~s^{-1}$, implying that the disc's outer edge velocity is $\sim 600 \rm ~km~s^{-1}$, and the emission-line wings indicate inner disc velocities of $\sim 2200 \rm ~km~s^{-1}$.  By comparing the double-peaked profile with the model of \cite{Horne1986}, the sharp peaks are characteristic of an optically thick disc and the deep central reversal suggests an intermediate to high inclination.  

\begin{figure}
\centering
\includegraphics[width=0.35\textwidth]{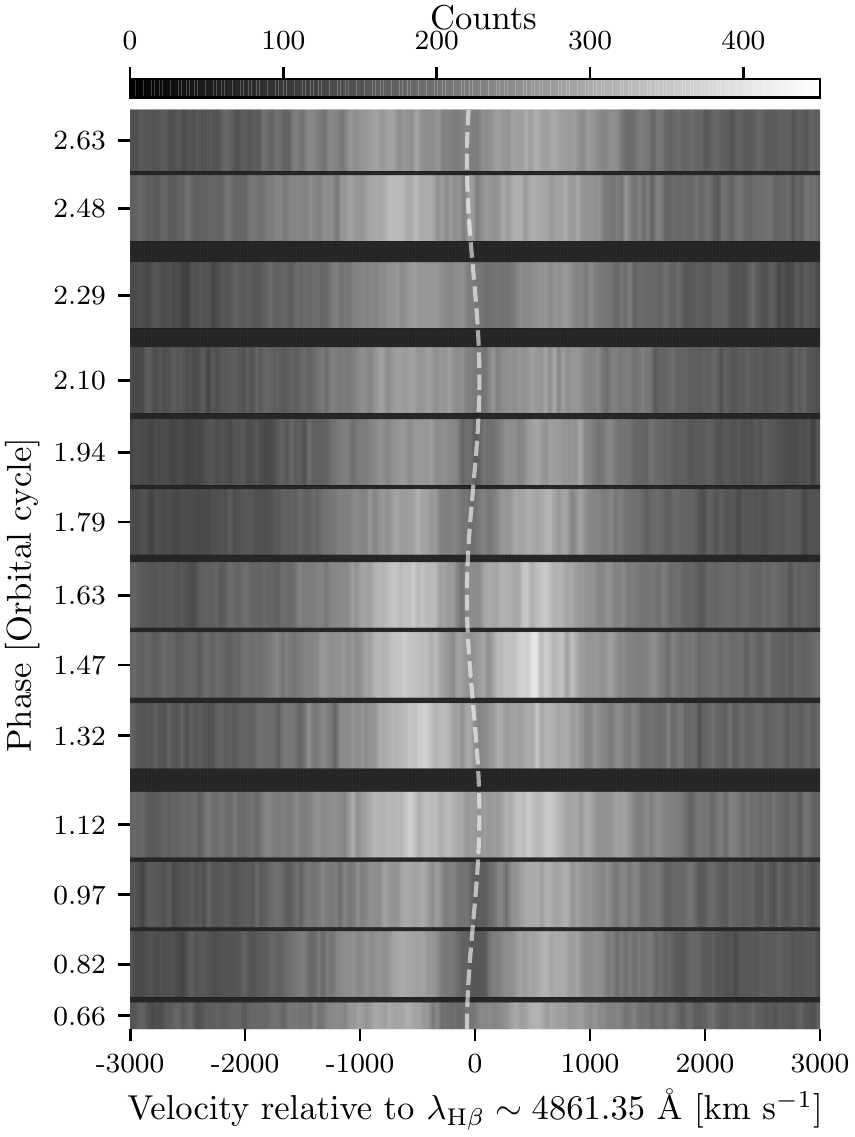}
\caption {H$\beta$ profiles of AR Pic on 26 Feb 2017 plotted as a grey-scale against orbital phase (using the known ephemeris). The dashed line is a sine fit to the radial velocity curve and illustrates the S-wave. An orbital change in overall line intensity correlates with the observer's viewing angle of the disc. The emission is weakest at phase 1.94 where the disc is at superior conjunction, and strongest at phase 1.47 when the disc is closest to the observer.}
\label{arpic_colormap}
\end{figure}

\begin{figure}
\centering
\includegraphics[width=0.45\textwidth]{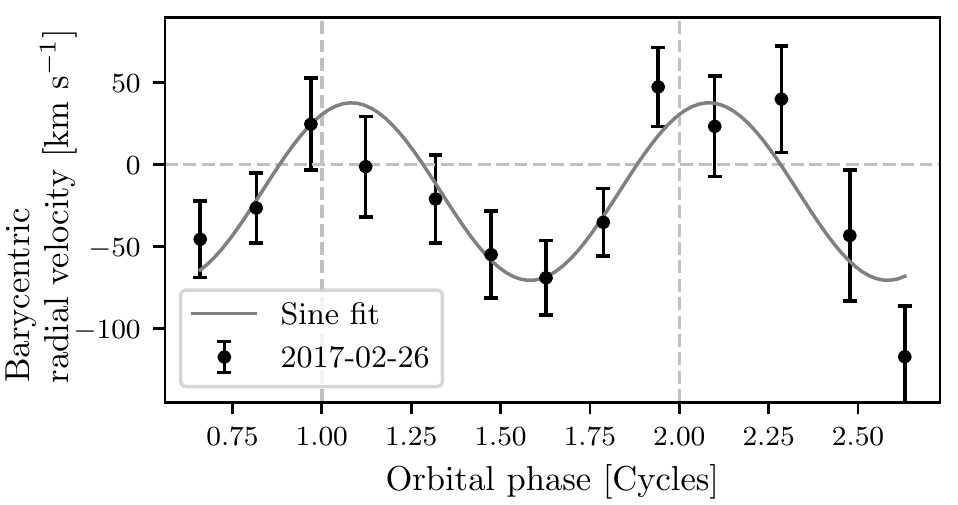}
\caption {H$\beta$ radial velocity curve of AR Pic. A sine fit gives a systemic velocity of $\gamma = -16\pm7 \rm ~km~s^{-1}$ and a semi-amplitude of $K_1 = 54\pm10 \rm ~km~s^{-1}$. Error bars show $\pm 1\sigma$.}
\label{arpic_rv}
\end{figure}

The H$\beta$ lines in the 26 February spectra are also plotted as a grey-scale in Fig.~\ref{arpic_colormap}, and an S-wave is clearly visible within the double-peaked profiles, likely due to the motion of the WD, on which the disc is centred. To measure $K_1$, the method of \citet{Shafter1983} was first applied, which entails a Gaussian function fitted through the wings of the double-peaked H$\beta$ line to determine the central wavelength. However, this produced an inconclusive radial velocity curve. Alternatively, two SplitGaussian functions were fitted to the double-peaked profiles from which average central wavelengths were calculated. The H$\beta$ barycentric-corrected radial velocity curve is shown in Fig.~\ref{arpic_rv}, and a sine function was fitted (gray curve in plot), giving a systemic velocity of $\gamma = -16\pm7 \rm ~km~s^{-1}$ and a semi-amplitude of $K_1 = 54\pm10 \rm ~km~s^{-1}$. The dashed line in Fig.~\ref{arpic_colormap} is the sine function fitted to the radial velocity curve and used to illustrate the S-wave.

\subsection{QW Ser}

The 2003--17 light curve of QW Ser (Fig.~\ref{longterm_lc}, middle panel) gives an average $V_{\rm q} \sim 18$ and reveals at least 24 outbursts with typical amplitude of $\sim 5$ mag. Using the same outburst criteria as mentioned in Section~\ref{arpic_section}, 6 were normal outbursts and 9 were superoutbursts. The low observing cadence prevented us from identifying them all. A duty cycle of $\sim 30 - 60$ days and $P_{\rm sc} \sim 270 - 330$ days were determined, the latter longer than the $P_{\rm sc} \sim 220 - 270$ days determined by \cite{Nogami2004a}, indicating a decrease in $\dot{M_2}$.  

We observed QW Ser in August 2016 just before and during an outburst, and followed it into quiescence; in 2017 it was observed only in quiescence.  The first four runs of photometry and spectroscopy in 2016 were conducted simultaneously and covered the pre-outburst (4 August), outburst (8 August) and decline (9 -- 10 August), making these the first outburst spectra to be presented for QW Ser. 

\begin{figure}
 \centering
 \includegraphics[width=0.47\textwidth]{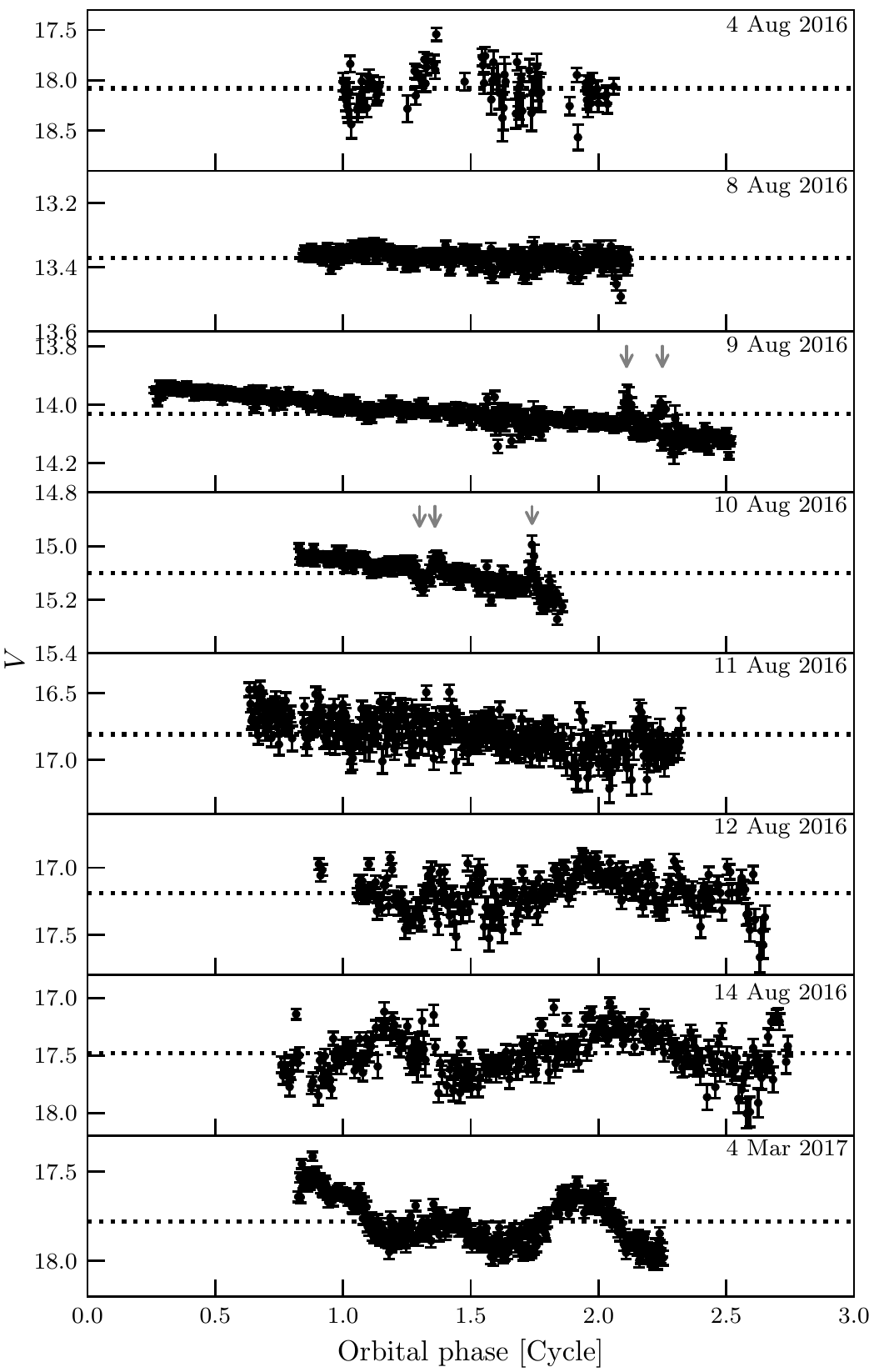}
 \caption{The light curves of QW Ser pre-outburst (4 Aug 2016), during a normal outburst (8 Aug 2016), declining (9 -- 12 and 14 Aug 2016) and in quiescence (4 Mar 2017). Phase zero is arbitrary, and is set to the first data point on 4 Aug.}
 \label{qwser_phase}
\end{figure}

Our photometry of QW Ser (Fig.~\ref{qwser_phase}) exploits the known $P_{\rm orb}$ ($ = 1.79$ hr), but uses an arbitrary phase zero of the first data point of 4 August.  The pre-outburst light curve (4 Aug) is noisy, as QW Ser is then very faint ($V \sim 18$), but does display variations similar to those seen later in the outburst. On 8 August, it was $\sim 4.5$ magnitudes brighter, likely close to outburst maximum, given the typical amplitudes seen.  No periodic features and little variability are seen on this or the subsequent two nights, indicating it must be a normal outburst, although a few fast variations were seen (marked with gray arrows), the nature of which is unclear. 

\begin{figure}
 \centering
 \includegraphics[width=0.475\textwidth]{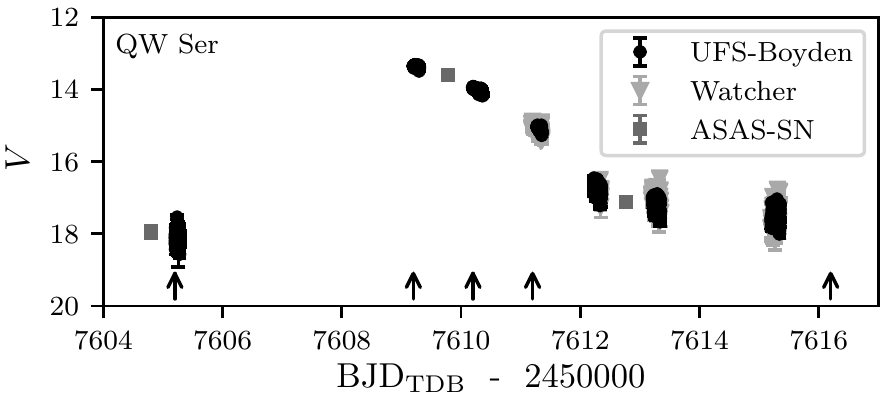}
 \caption{Overall light curve of QW Ser on 4 -- 14 Aug 2016, supplemented by ASAS-SN. It increased by $\sim 4.5$ magnitudes during outburst and returned to quiescence within a week. Black arrows indicate dates of spectroscopy.}
 \label{qwser_outburst}
\end{figure}

On 9 August QW Ser was fading at $\sim 0.06 \rm~mag~hr^{-1}$ and reached quiescence within a week (Fig.~\ref{qwser_outburst}) at a mean rate of $0.68 \rm~mag~day^{-1}$. The orbital modulation became visible on August 12 and 14, as QW Ser approached quiescence. The lack of any eclipses suggests an intermediate inclination of $i<75^{\circ}$. 

Two humps of unequal amplitude were detected during one quiescent orbital cycle (4 March).  A Lomb-Scargle periodogram indicated a period of $P_{\rm LS} \sim 0.92\pm0.17$ hr, i.e. half of $\sim P_{\rm orb}$. The humps suggest an observable hot spot, and further photometry in quiescence is  required to establish an ephemeris.

\begin{figure*}
 \centering
 \includegraphics[width=0.8\textwidth]{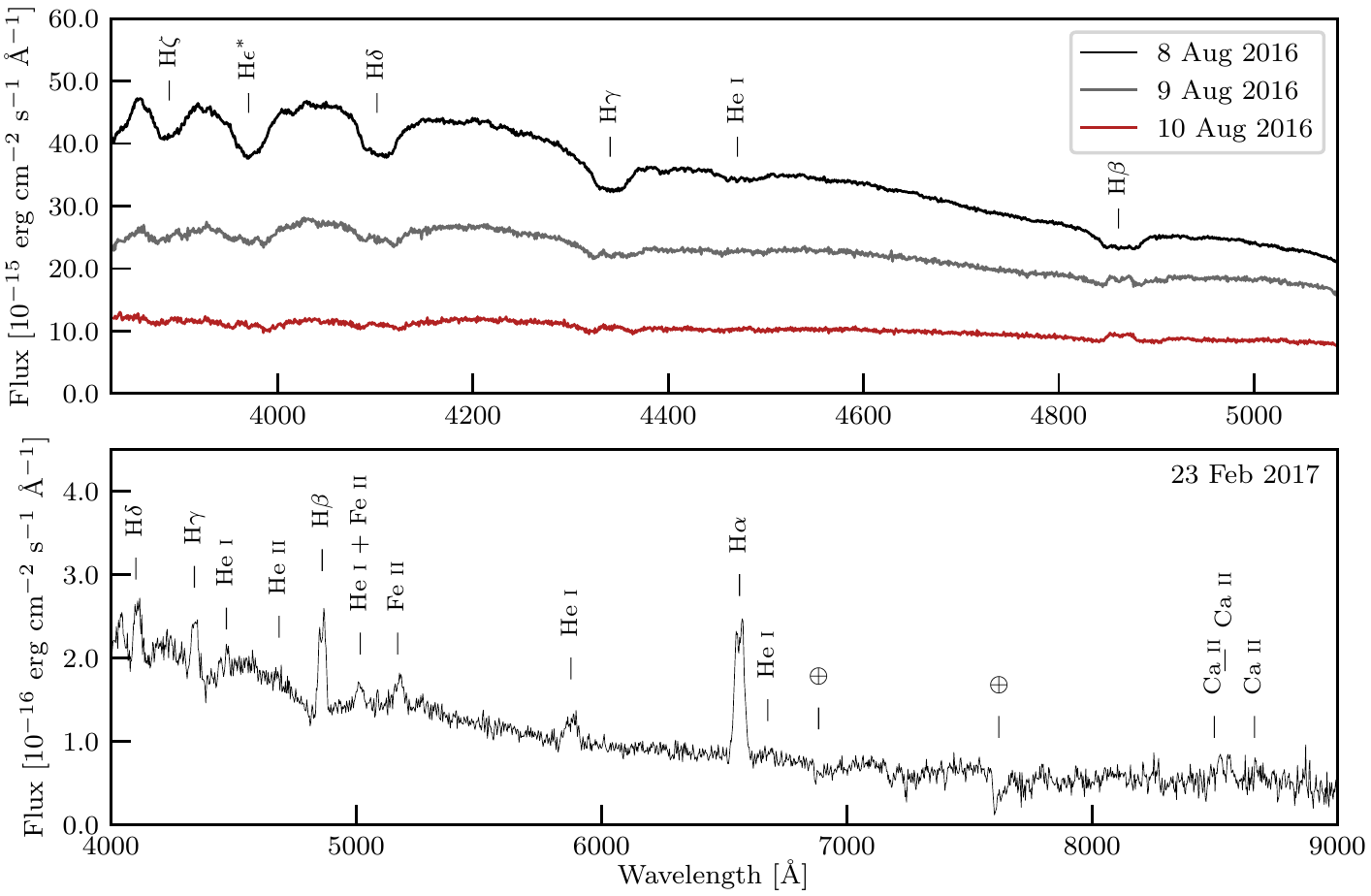}
 \caption{\textit{Top panel}: Spectra of QW Ser obtained during a normal outburst (8 Aug 2016) and while fading back to quiescence (9 and 10 Aug 2016). The Balmer absorption features are typical of an optically thick disc during outburst. As the system returned to quiescence, the disc cooled and became optically thin, allowing the emission lines to return. The spectra of 9 and 10 Aug are multiplied by a factor of 2 to amplify features. \textit{Bottom panel}: Spectrum of QW Ser in quiescence (23 Feb 2017). The double-peaked emission lines are evidence of an accretion disc.}
 \label{qwser_spec1}
\end{figure*}

Our quiescent spectra of QW Ser on  23 February 2017 (Fig.~\ref{qwser_spec1}, bottom panel, is the best S/N spectrum obtained) show a typical DN with H and He{\sc~i} emission lines. The double-peaked emission profiles show the presence of an accretion disc. No secondary star features were detected; however, the quiescent spectrum does have broad absorption wings flanking H$\beta$, H$\gamma$ and H$\delta$ emission, likely originating from the hot WD. 

\begin{figure}
 \centering
 \includegraphics[width=0.48\textwidth]{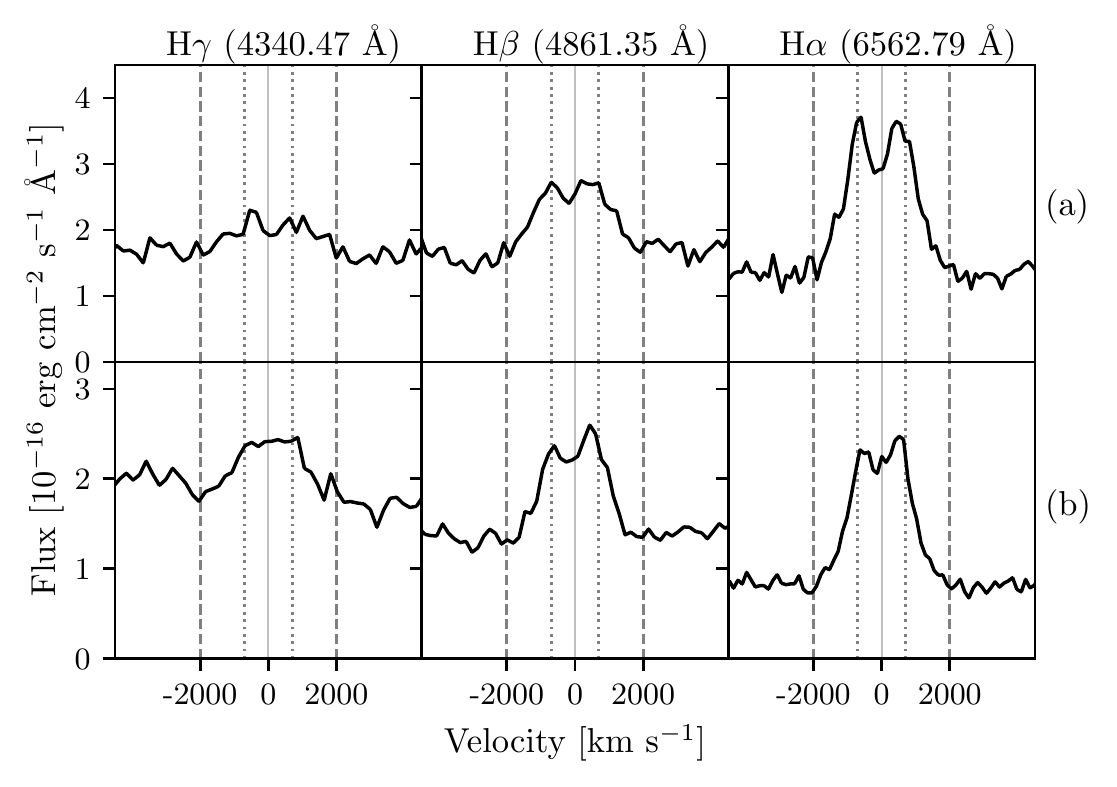}
 \caption{Balmer lines in the quiescent spectra of QW Ser obtained on (a) 4 Aug 2016 and (b) 23 Feb 2017. Dashed and dotted lines are as in Fig.~\ref{arpic_beta}, and give a peak separation of $\sim 1400 \rm ~km~s^{-1}$.}
 \label{qwser_quies_sub}
\end{figure}

The quiescent Balmer profiles from 4 August 2016 and 23 February 2017 (Fig.~\ref{qwser_quies_sub}) indicate inner disc velocities of $\sim 2000 \rm ~km~s^{-1}$, based on the line wings. The peak separation of $\sim 1400 \rm ~km~s^{-1}$ imply outer disc velocities of $\sim 700 \rm ~km~s^{-1}$. The peak separation and $\vee$-shaped central reversal in the pre-outburst spectrum (Fig.~\ref{qwser_quies_sub} (a)) is consistent with an optically thin accretion disc viewed at an intermediate inclination. The asymmetrical peaks in the H$\alpha$ and H$\beta$ lines in Fig.~\ref{qwser_quies_sub} (b), are evidence of an observable hot spot, as implied in our photometry.    

The absorption lines visible in the outburst spectrum (Fig.~\ref{qwser_spec1}, 8 Aug 2016) are typical of a hot, optically thick disc.  As QW Ser returned to quiescence, the disc cooled and became optically thin, allowing for the emission lines to return, as seen in Fig.~\ref{qwser_spec1}, 9--10 Aug 2016. The parameters we derived for QW Ser match previously observed values.


\subsection{V521 Peg}

The composite light curve of V521 Peg (Jul 2003 -- Nov 2019, Fig.~\ref{longterm_lc} bottom panel) shows an average $V_{\rm q} \sim 17.2$, in agreement with \citet{Coppejans2016}. Seventeen outbursts were detected with an average outburst amplitude of $\sim 5$ mag. To determine whether they were normal or superoutbursts, the same outburst criteria in Section~\ref{arpic_section} were applied and revealed 2 normal outbursts and 7 superoutbursts, but the nature of the remainder could not be identified due to the low observing cadence. Based on the superoutburst time intervals, they recur on a timescale of $P_{\rm sc} \sim 320 - 350$ days, longer than the $P_{\rm sc} \sim 300$ day supercycle determined by \citet{Patterson2011}, and indicate a decrease in $\dot{M_2}$. A normal outburst duty cycle could not be determined. Despite the fact that the nature of not all the outbursts could be determined, the lack of normal outbursts supports the ``low-activity'' observation of \cite{Aung2006}.

\begin{figure}
 \centering
 \includegraphics[width=\columnwidth]{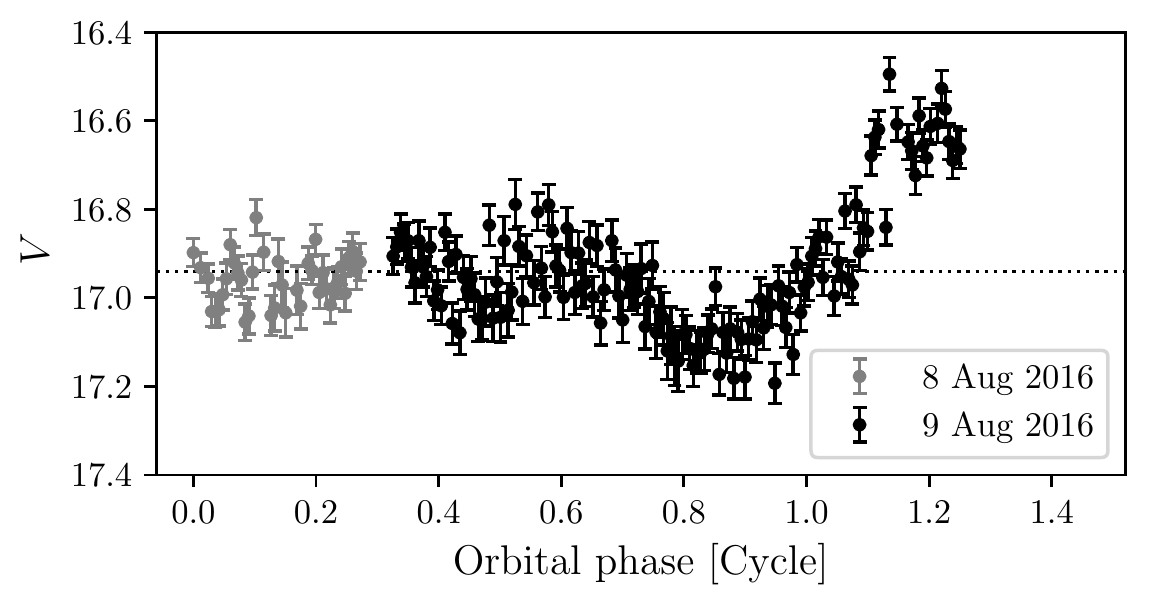}
 \caption{The quiescent light curves of V521 Peg on 8 Aug (gray) and 9 Aug (black) 2016.}
 \label{v521peg_quies}
\end{figure}

Our quiescent photometry of V521 Peg  (8--9 Aug 2016) revealed no eclipses (Fig.~\ref{v521peg_quies}), but the double-hump structure on 9 August is indicative of a hot spot. Although not as well defined as that of RG05, the light curve does resemble the same double-hump structure with two peaks of unequal amplitude. 

\begin{figure*}
 \centering
 \includegraphics[width=0.8\textwidth]{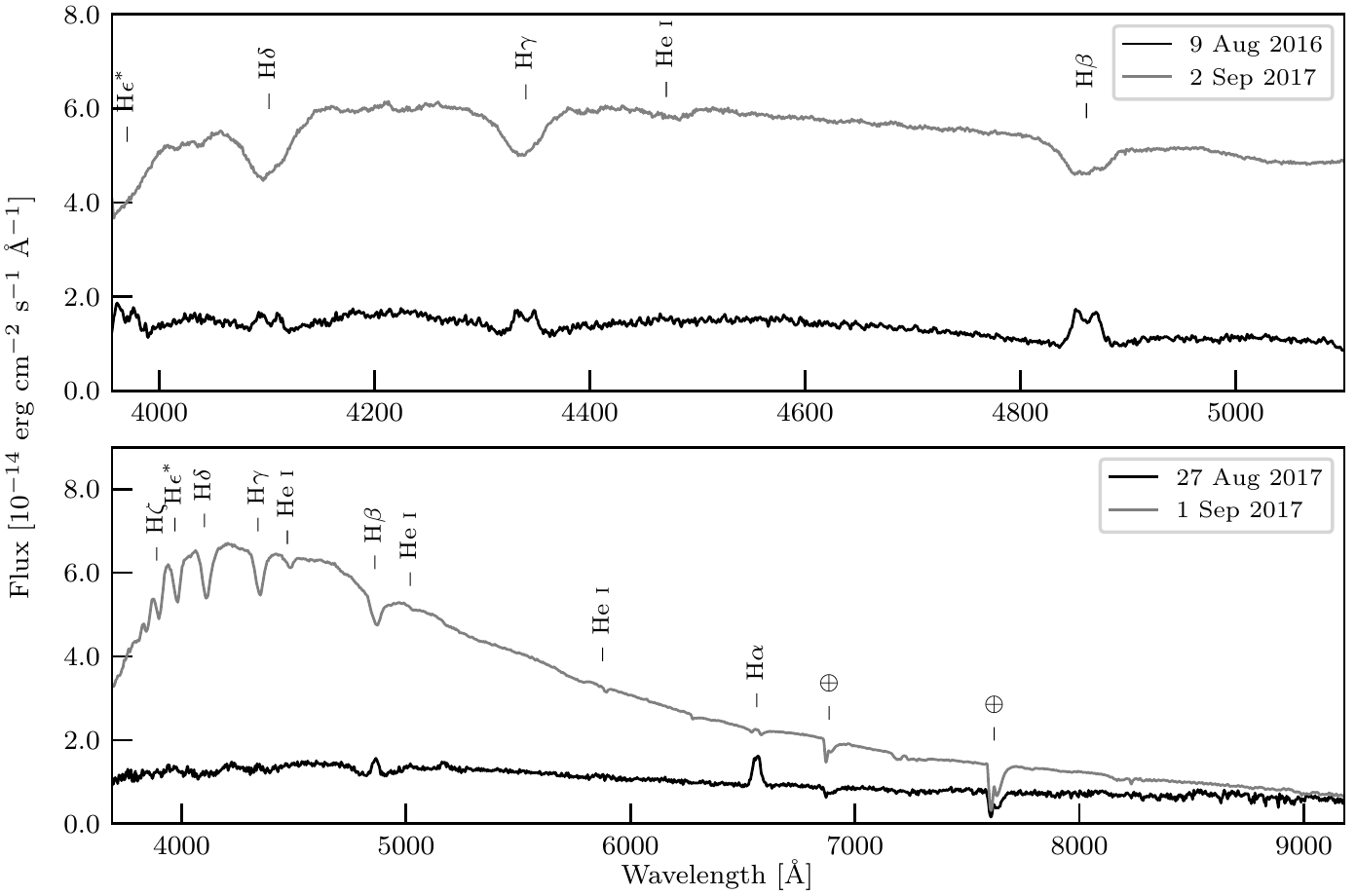}
 \caption{\textit{Top panel}: Higher resolution spectra of V521 Peg obtained during quiescence (9 Aug 2016) and a superoutburst (2 Sep 2017). \textit{Bottom panel}: Spectra of V521 Peg obtained during a pre-cursor outburst (27 Aug 2017 - note that this was observed at high airmass and could not be reliably fluxed) and a superoutburst (1 Sep 2017). The spectra of 9 Aug 2016 and 27 Aug 2017 are multiplied by a factor of 20 to amplify features. The double-peaked emission lines (9 Aug) are evidence of an accretion disc. The absorption features in the superoutburst spectra indicate that the disc is optically thick during superoutburst. H$\alpha$ is almost undetectable in the 1 Sep spectrum as it consists of an absorption line with a superimposed emission component.}
 \label{v521peg_spec}
\end{figure*}

The 9 August quiescent spectrum of V521 Peg (Fig.~\ref{v521peg_spec}, top panel) displayed double-peaked Balmer emission lines, indicating an accretion disc viewed at moderate inclination. The Balmer emission lines were flanked by broad absorption wings which are attributed to the pressure-broadened features originating in the WD photosphere. No late-type features from the secondary were detected. 

\begin{figure}
 \centering
 \includegraphics[width=\columnwidth]{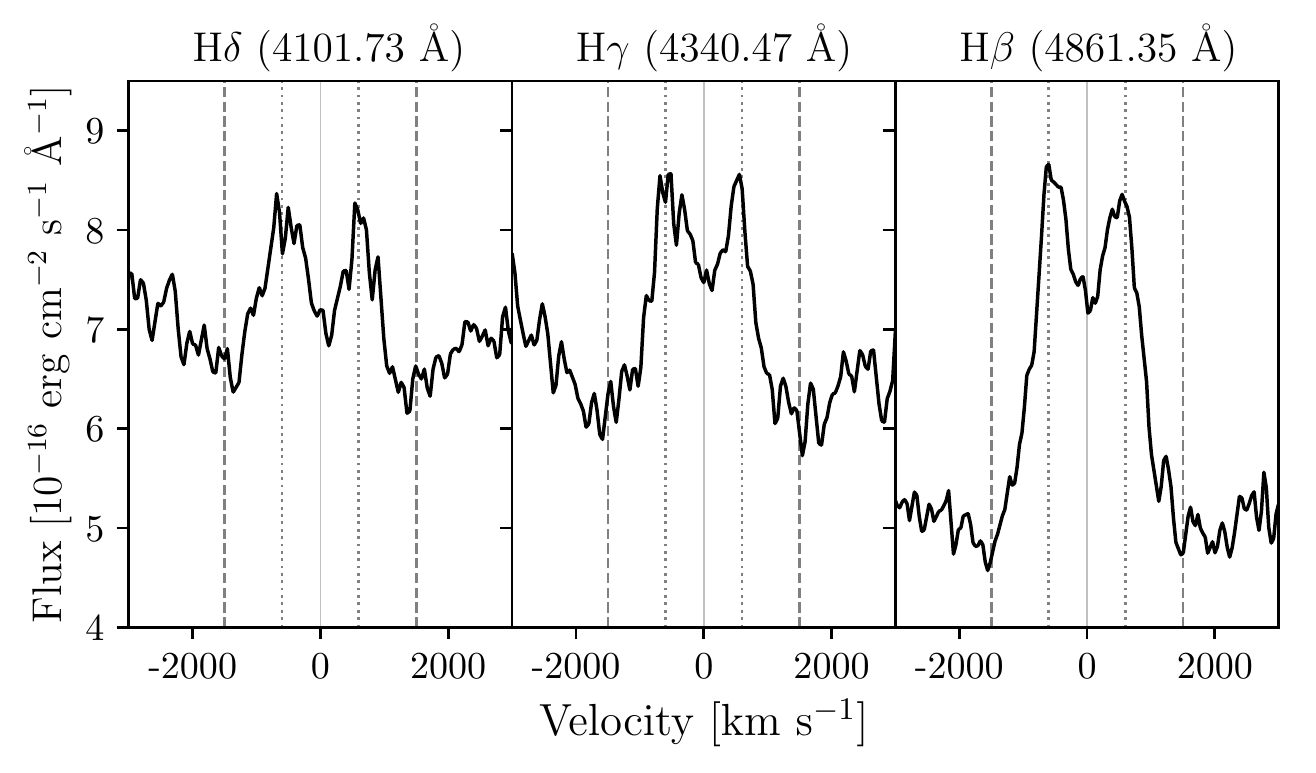}
 \caption{Balmer lines in the 9 August 2016 quiescent spectrum of V521 Peg.  Dashed and dotted lines are as in Fig.~\ref{arpic_beta}, and give a peak separation of $\sim 1200 \rm ~km~s^{-1}$, and velocity close to the WD of $\sim 1500 \rm ~km~s^{-1}$.}
 \label{v521peg_quies_sub}
\end{figure}

Fig.~\ref{v521peg_quies_sub} shows the Balmer lines in the 9 August 2016 quiescent spectrum, where the peaks are slightly asymmetrical, providing evidence of a hot spot.  The peak separation of $\sim 1200 \rm ~km~s^{-1}$ implies an outer disc edge velocity of $\sim 600 \rm ~km~s^{-1}$, in agreement with that of RG05 ($\sim 660 \rm ~km~s^{-1} $). The inner disc velocities close to the WD of $\sim 1500 \rm ~km~s^{-1}$ were estimated from the wing limits, but this should be considered a lower limit due to the difficulty in defining the edges of these features. The $\vee$-shaped central reversal at the peak is typical of an optically thick accretion disc viewed at an intermediate inclination \citep{Horne1986}.  

During observations on 1 September 2017 we found V521 Peg to be in outburst, and therefore undertook simultaneous photometry and spectroscopy at Boyden and SAAO for the following three nights (2--4 Sep, see Tables~\ref{spec_obs} and \ref{phot_obs}).  Coincidentally the AAVSO community and ASAS-SN also observed the source pre-outburst and during the superoutburst, completing the dataset used here.

\begin{figure*}
 \centering
 \includegraphics[width=0.9\textwidth]{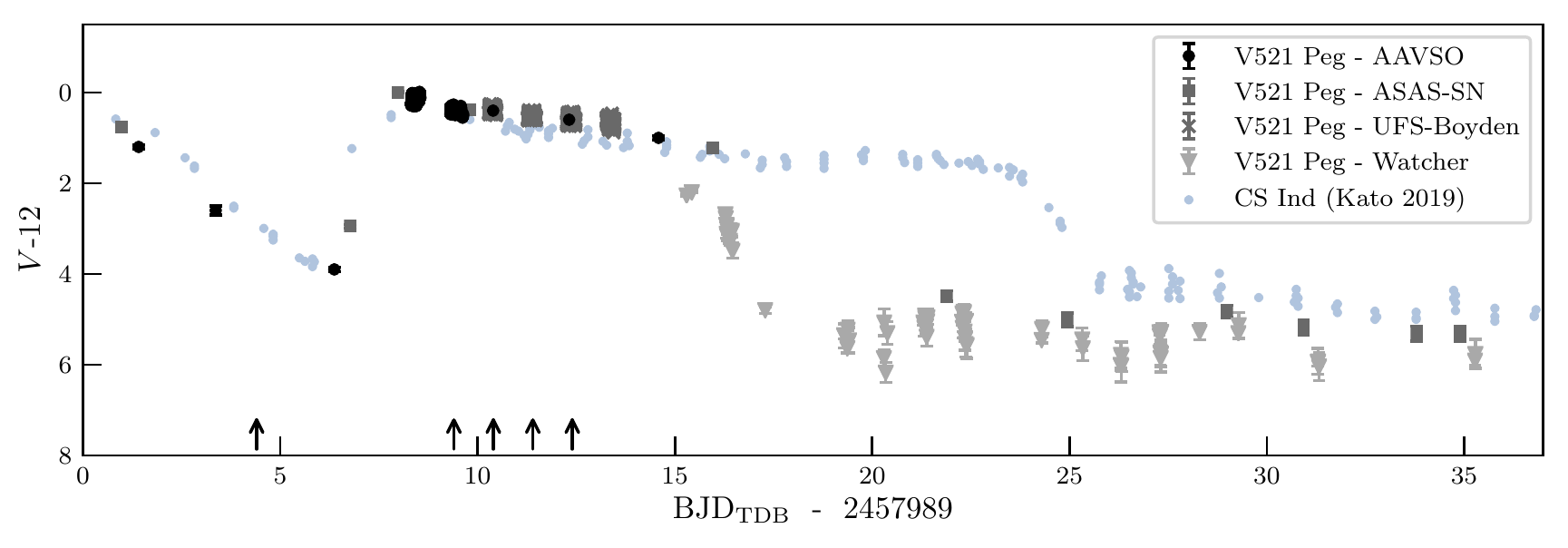}
 \caption{The overall light curve of V521 Peg for 24 Aug -- 27 Sep 2017, showing the precursor outburst, the superoutburst, and return to quiescence within 10 days. The precursor and superoutburst of CS Ind in Nov 2018 (light blue data points; \citealp[adapted from][Fig.~1]{Kato2019}) are plotted as a comparison. }
 \label{v521peg_C}
\end{figure*}

The overall light curve (Fig.~\ref{v521peg_C}) reveals that V521 Peg was $\sim 5$ mag brighter in superoutburst than in quiescence. During the plateau phase it declined at a rate of $0.1 \rm ~mag~day^{-1}$ and more rapidly thereafter at $0.9 \rm ~mag~day^{-1}$, returning to quiescence within 10 days. This superoutburst duration is shorter than that observed in 2003 by RG05, where the plateau phase alone was longer than 10 days. 

Fig.~\ref{v521peg_C} shows that, in fact, V521 Peg was declining from a normal outburst that led directly into the superoutburst. These precursor outbursts usually manifest as a small shoulder that evolves quickly into the superoutburst \citep{Cannizzo2010}, however, the detected dip in this case is unusual as it had declined almost 4 mag within a week before the onset of superoutburst. This is actually similar to well-separated precursor outbursts observed for other standard SU UMa systems, e.g. VW Hyi \citep{Bateson1977}, V1504 Cyg and V344 Lyr \citep{Osaki2014}, and CS Ind \citep{Kato2019}, with the latter included in Fig.~\ref{v521peg_C} (light blue data points) to demonstrate their similarity. \cite{Osaki2014} proposed that the well-separated precursor outbursts are due to the slow growth of the tidal instability that can survive quiescence and eventually trigger the superoutburst. They also pointed out that the interval between well-separated precursor outbursts and the superoutbursts (i.e. time between maximum magnitudes) for high-$\dot{M_2}$ systems is $\sim ~4 - 5$ days, while for low-$\dot{M_2}$ systems it is $10 - 11$ days. This interval for V521 Peg is a minimum of $\sim 8$ days since the date of maximum magnitude of the precursor outburst is unknown, hence indicating that V521 Peg is a low $\dot{M_2}$-system.

\begin{figure}
 \centering
 \includegraphics[width=\columnwidth]{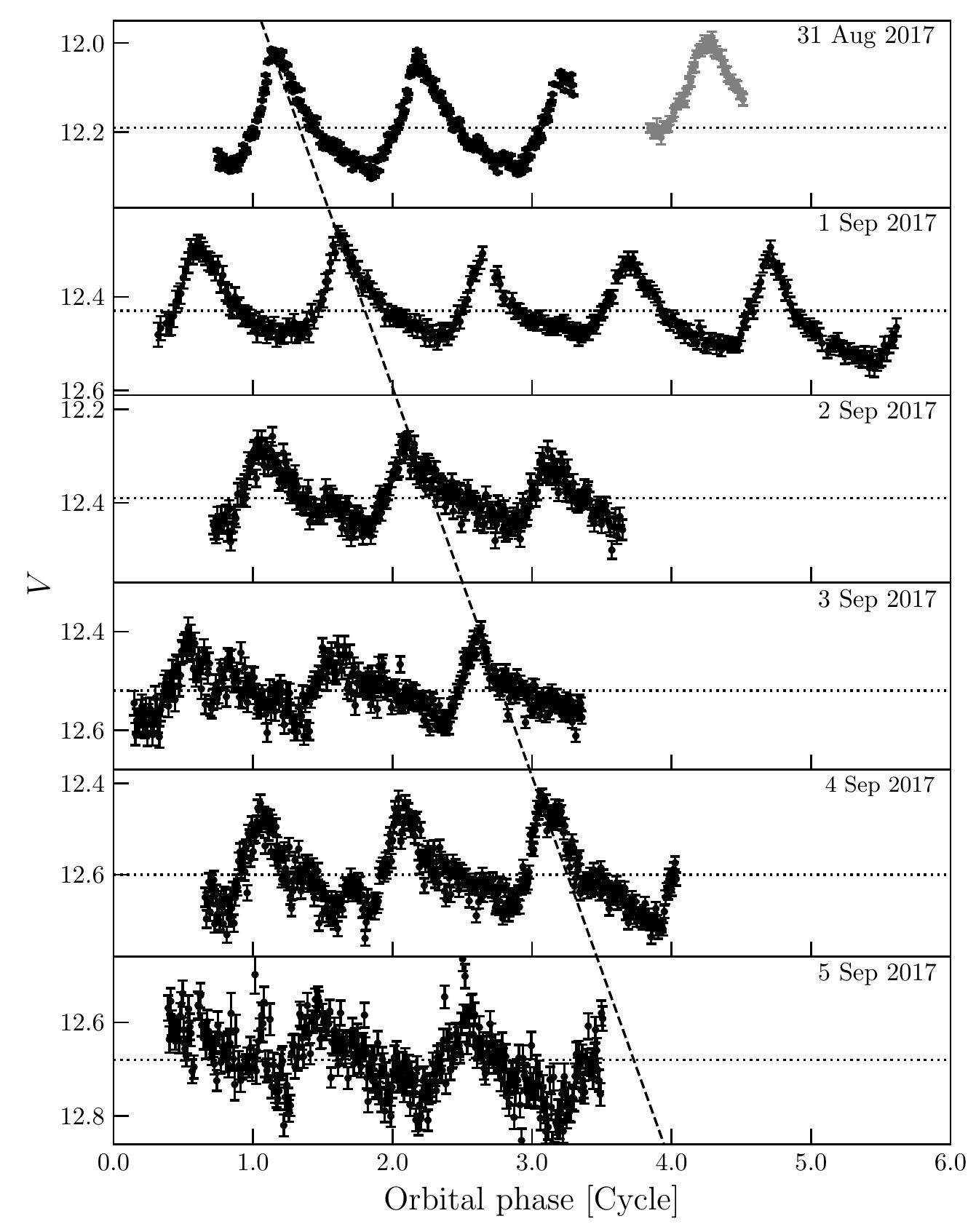}
 \caption{The AAVSO (31 Aug -- 1 Sep) and Boyden (2 -- 5 Sep) superoutburst V-band light curves of V521 Peg plotted against orbital phase using $P_{\rm orb} = 1.44$ hr. Black points used a C-filter and were reduced to $V$, while gray points used a $V$-filter. The dashed line highlights the superhump drift relative to orbital phase, as $P_{\rm sup}$ is a few percent longer than $P_{\rm orb}$. The dotted lines represent the average magnitude for each run.}
 \label{v521peg_out}
\end{figure}

The presence of superhumps during the plateau phase (Fig.~\ref{v521peg_out}) confirmed that it was indeed a superoutburst, and they drifted as expected when plotted against orbital phase using the known $P_{\rm orb} \sim 1.44$ hr. All data points were reduced to $V$, but given that AAVSO observers use a range of cameras with different responses, there will likely be small systematic differences present. The Boyden data were binned in steps of two 10 sec frames in order to smooth the light curves. A maximum superhump amplitude of $\sim 0.3$ mag was observed on 31 August and subsequently decreased.   

\begin{figure}
 \centering
 \includegraphics[width=\columnwidth]{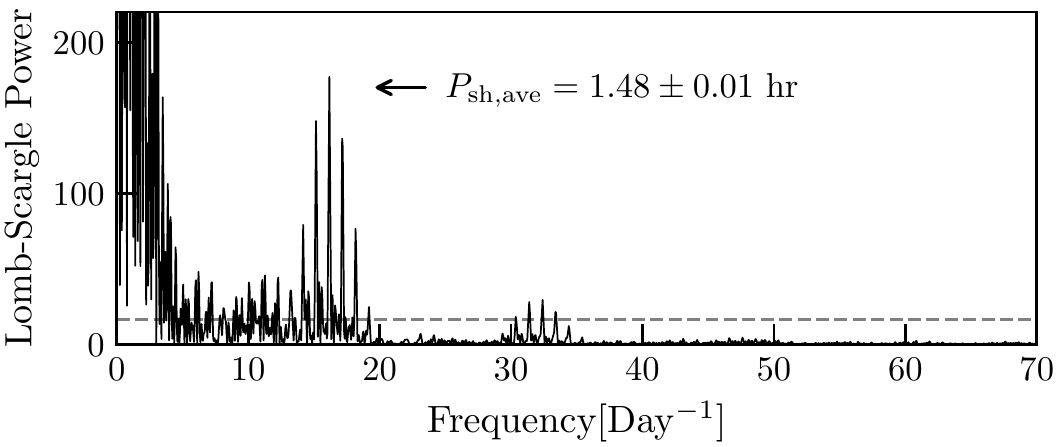}
 \caption{The Lomb-Scargle periodogram of the superoutburst light curves shown in Fig.~\ref{v521peg_out}. An average superhump period of $P_{\rm sh,ave} = 1.48 \pm 0.01 \rm ~hr$ was determined. The dashed lines represent the $3\sigma$ significance level for white noise.}
 \label{v521peg_superhump_ls}
\end{figure}

A Lomb-Scargle (LS) periodogram of these data\footnote{Produced using the software package Starlink {\small PERIOD} (Version 5.0-2) \citep{Dhillon2001}} is shown in Fig.~\ref{v521peg_superhump_ls}, and gave an average $P_{\rm sh} = 1.48\pm0.01$ hr, i.e. a period excess of $\varepsilon = (P_{\rm sh} - P_{\rm orb})/P_{\rm orb} \approx 0.03$.

Our superoutburst spectra of V521 Peg (Fig.~\ref{v521peg_spec}, 1--2 Sep 2017), the first to be obtained for this source, was significantly bluer than in quiescence and showed the presence of broad H and He{\sc~i} absorption. However, H$\alpha$ was almost undetectable (Fig.~\ref{v521peg_spec}, bottom panel). These spectra are similar to those of WZ Sge (2001 superoutburst, \citealp{Nogami2004b}) and GW Lib (2007 superoutburst, \citealp{Hiroi2009}) and typical of an optically thick accretion disc. H$\alpha$, which appeared very faint, was in effect an absorption line filled in by emission, the latter arising from the hot inner disc/WD irradiating the outer disc \citep{Hiroi2009}.  We also have spectra from 3 -- 4 September, but these were not included as they are similar to those in Fig.~\ref{v521peg_spec}, except for a decline in flux. These spectra are available as supplementary material.

\subsection{Orbital period - Supercycle relation}

\begin{figure}
 \centering
 \includegraphics[width=\columnwidth]{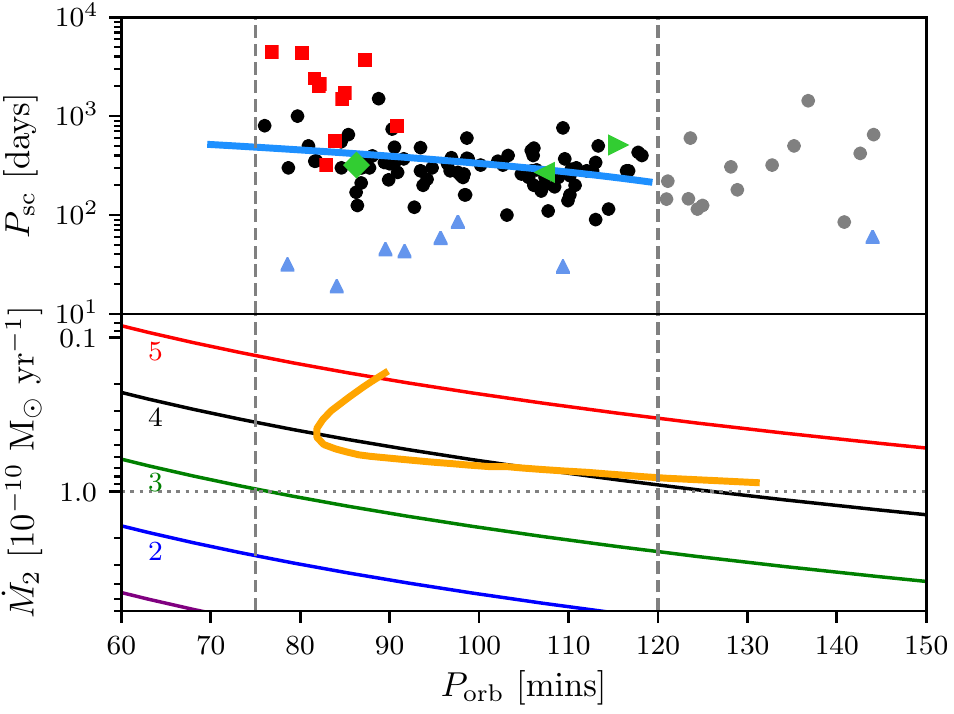}
 \caption{\textit{Top panel}: Orbital period ($P_{\rm orb}$) - Supercycle ($P_{\rm sc}$) relation for SU UMa (black and gray circles), ER UMa (pale blue triangles) and WZ Sge-type (red squares) systems. Values obtained from \citealp{Hypka2016} and include AR Pic (green triangle pointing right), QW Ser (green triangle pointing left) and V521 Peg (green diamond) with updated $P_{\rm sc}$. The blue line is the linear fit to the SU UMa-type systems with $P_{\rm orb} <$ 120 min (black circles), and indicates an increase in $P_{\rm sc}$ as $P_{\rm orb}$ decreases. \textit{Bottom panel}: The orange line is the mass transfer rate ($\dot{M_2}$) -- Orbital period ($P_{\rm orb}$) relation of the revised evolution model for DNe given by \citet[Fig.~19]{Knigge2011b}. The red, black, green and blue lines represent the $\dot{M}_2$ -- $P_{\rm orb}$ relation given in Eq.~\ref{mass_rate} for different values of $\alpha$ ($2 \leq \alpha \geq 5$). }
 \label{sc_orbital}
\end{figure}

Having determined the times between consecutive superoutbursts ($P_{\rm sc}$) using the archival data for our targets, it is interesting to examine the relation between $P_{\rm sc}$ and $P_{\rm orb}$ for those SU UMa-systems and related objects where these parameters are known. By combining catalogue data \citep{Hypka2016} with our composite light curves (Fig.~\ref{longterm_lc}), we produced Fig.~\ref{sc_orbital} (top panel) where we have also included the WZ Sge and ER UMa systems. Note that the three subclasses inhabit different, clearly defined regions of this plot. The WZ-type (red squares) have extremely rare superoutbursts ($P_{\rm sc} > 800$ days) and are concentrated close to the `period minimum` of $\sim$80 mins.  In comparison, the ER-type (blue triangles) have extremely frequent super and normal outbursts ($P_{\rm sc} \sim 20 - 100$ days) and yet are spread over a much wider range of $P_{\rm orb}$.  The SU-type (black and gray circles) occupy the ``middle ground'' between these types. A linear function was fitted to the 81 SU UMa data points below the `period gap' ($P_{\rm orb} < 2$ hrs), utilizing Python scipy.stats.linregress to calculate a linear least-squares regression, which gave
\begin{equation}
   P_{\rm sc} = -(4.9 \pm 1.6)~P_{\rm orb} + (811.9 \pm 157.4)
\end{equation}
By performing a t-test on our data we find that it rejects the null hypothesis slope of zero at a $3\sigma$ confidence level (99.7\%); we cannot undertake a standard goodness-of-fit analysis because the catalogue data do not include uncertainties in the $P_{\rm sc}$ values. DT~Pyx ($P_{\rm orb} = 88.78$ min, $P_{\rm sc} = 1500$ days) was excluded from our fit, as while it is classified as a SU-type, its $P_{\rm sc}$ value is above the SU-type range, and puts it much closer to a WZ-type. It also has a quiescent $\rm M_V = +11.4$ which is comparable with WZ-types and fainter than the typical SU-type (e.g. \citealp[][Fig.~3]{Warner1987}; \citealp[][Fig.~9.8]{Warner1995}). Equation (2) is plotted as the blue line in Fig.~\ref{sc_orbital} (top panel) and shows that SU-type sources tend towards increasing $P_{\rm sc}$ at shorter $P_{\rm orb}$. Our $P_{\rm sc} - P_{\rm orb}$ plot is also consistent with the recent results of \citep[][Fig.~1]{Vogt2021} who did not perform any fits to their data.

Angular momentum loss in close binary systems causes most CVs, after they emerge from the `period gap', to evolve down to a `period minimum' of $P_{\rm orb,min} \sim $70 -- 80 min before they start to evolve back to longer orbital periods (see \citealp{Rappaport2017}, and references therein).  Detailed calculations of CV evolution (e.g. \citealp{Knigge2011b}) show that, after a spike in the mass transfer rate on emerging from the period gap, $\dot{M_2}$ quickly settles to $\sim 10^{-10} {\rm ~M_{\sun}~yr^{-1}}$ and gradually declines as the CV evolves to $P_{\rm orb,min}$.  It has been suggested (e.g. \citealp{Kato2013ERUMa}, who produce a similar plot to our Fig.~\ref{sc_orbital} top panel, but using $P_{\rm sh}$ instead of $P_{\rm orb}$), that this decline in $\dot{M_2}$ will reduce the rate of accumulation of matter in the disc, and hence increase the intervals between superoutbursts. We included the mass transfer rate ($\dot{M_2}$) -- orbital period ($P_{\rm orb}$) relation of the revised evolution model for DNe given by \citet[Fig.~19]{Knigge2011b} in the bottom panel of our Fig.~\ref{sc_orbital} (orange line). This revised model demonstrates that the mass transfer rate decreases as the system evolves to shorter orbital periods, hence resulting in a reduction of the rate of accumulation of matter in the disc, and leading to an increase in $P_{\rm sc}$.

The clear distinction between the SU-type and WZ/ER subclasses in terms of their different supercycle times close to $P_{\rm orb,min}$, indicates that the mass transfer rates are significantly different for each subclass.  Actually, the SU UMa systems with the shortest $P_{\rm orb}$ appear to blend very smoothly into the WZ Sge systems, further strengthening their combination into a single class (e.g. \citealp{Kato2015}).  What is not clear is the mechanism that is driving $\dot{M_2}$ to be so much higher in the ER UMa systems. 

In an effort to determine the distinguishing factor between the subclasses, we looked at the distance between the secondary's photosphere and the L1 funnel through which material flows, as a contributing factor. In this context, the mass transfer rate can be determined by 
\begin{equation}
    \label{mass_rate}
    -\dot{M_2} \approx \frac{1}{4~\pi}\rho_{\rm L1}c_{\rm s}^3 P^2_{\rm orb},
\end{equation}
where $\rho_{\rm L1}$ is the average density in the L1 region given by
\begin{equation}
    \rho_{\rm L1} = \frac{1}{\sqrt{\rm e}} \rho_{\rm phot} {\rm exp}\frac{-(R_2-R_{\ast})}{H_{\rm p}},
\end{equation}
 with $\rho_{\rm phot}$ representing the photospheric density of a late-type star, and is $\sim 10^{-6} {\rm ~g~cm^{-3}}$ (\citealp[e.g.][p.~34]{Warner1995}; \citealp[][p.~352]{Frank2002}). Applying the methodology of  \cite{Meintjes2004} for late-type dwarf stars, the mass transfer rates were calculated using different scale height factors ($\alpha$) determined by  
\begin{equation}
    \alpha = \frac{R_2-R_{\ast}}{H_{\rm p}},
\end{equation}
where $R_2$ is the secondary's Roche lobe radius, $R_{\ast}$ the photospheric radius, and $H_{\rm p} = c_{\rm s}^2 R_{\ast}^2/GM_2$ is the stellar scale height. Hence, a small distance between the photosphere and L1 will result in a small $\alpha$, whereas a large distance will give a large $\alpha$.

The mass transfer rates for various $\alpha$ values ($2 \leq \alpha \geq 5$) were plotted in Fig.~\ref{sc_orbital} (bottom panel) and all results indicate $\dot{M_2}$ decreases at shorter $P_{\rm orb}$. The plots indicate that small scaleheight factors ($\alpha \sim 2 - 3$) result in high $\dot{M_2}$, whereas the higher scaleheight factors ($\alpha \sim 4 - 5$) result in lower $\dot{M_2}$. It is possible that ER-types have scaleheight factors of $\alpha \sim 2 - 3$, distinguishing them from SU/WZ-types. The mass transfer rates for $\alpha \sim 4$ (typical for SU-types) compare well with the revised evolution model of \cite{Knigge2011b}. It has long been suspected that WZ-types are evolving through the `period minimum' \citep[e.g][p.~90]{Hellier2001} and becoming `period bouncers'. The grouping of these systems in Fig.~\ref{sc_orbital} (top panel) compares well with the lowest $\dot{M_2} < 0.4 \times 10^{-10} {\rm ~M_{\sun}~yr^{-1}} $ branch of the \cite{Knigge2011a} model (Fig.~\ref{sc_orbital}, bottom panel). This suggests that WZ-types are evolving from a scaleheight factor of $\alpha \sim 4$ towards $\alpha \sim 5$.

For systems that have reached $P_{\rm orb,min}$, the donor can no longer support H-burning and it starts to become degenerate (e.g. \citealp[p.~53]{Hellier2001}, and references therein). These stars have convective envelopes with a typical adiabatic index of $\zeta = -0.2$ \citep[e.g.][and references therein]{Meintjes2002}, and expands according to the M-R relationship for degenerate secondary stars, i.e. $R \propto M^{-\beta}$ ($\beta \sim 0.8$; e.g. \citealp{Hamada1961, Eracleous1996}) maintaining Roche contact to transfer mass at much lower rates. To conserve angular momentum, the system expands which leads to an expanding Roche lobe. These `post-bounce' donors are of much lower mass ($\leq 0.05$M$_\odot$) and so provide a much lower $\dot{M_2}$, making these CVs much fainter.

AR Pic, QW Ser and V521 Peg are included in Fig.~\ref{sc_orbital} (top panel) and their positions show that they are still evolving toward the `period minimum'. AR Pic and QW Ser will remain SU UMa sources for a long time to come, however, V521 Peg is within the region that overlaps with WZ Sge sources. The photometric and spectroscopic results of V521 Peg show that it has characteristics of both SU UMa (precursor outburst comparable with SU-types) and WZ Sge types (lack of normal outbursts, superoutburst spectrum similar to WZ Sge). This might support the argument that SU UMa and WZ Sge-types are a single class \citep[e.g.][]{Kato2015}, however, the different scaleheight factors for these subclasses can be a distinguishing factor and perhaps cast doubt on the single class argument. It is clear that V521 Peg is on the cusp of evolving into a WZ Sge-type system.

\section{Conclusions}

Long-term archival monitoring (with CRTS, AAVSO, ASAS-SN and ASAS-3) has allowed us to estimate their supercycle recurrence times ($P_{\rm sc}$) and we determined that $P_{\rm sc}$ has increased for QW Ser and V521 Peg, indicating a decreasing $\dot{M_2}$ in these systems. For AR Pic, $P_{\rm sc}$ was only better constrained and requires further long-term monitoring to determine if there is a change in $P_{\rm sc}$. Optical photometry and spectroscopy of our three target dwarf novae covering quiescence, normal and super-outbursts has shown the characteristic properties of SU UMa-systems. Although a change in $P_{\rm sc}$ was detected for QW Ser, it's outburst properties did not deviate from previous outbursts. V521 Peg, however, exhibited an uncommon well-separated precursor outburst that preceded a superoutburst, the latter not lasting as long as a previous recorded superoutburst.

We combined our $P_{\rm sc}$ results with the catalogued $P_{\rm sc}$ for all short-period dwarf novae, showing a possible relation between $P_{\rm orb}$ and  $P_{\rm sc}$, whereby $P_{\rm sc}$ increases as $P_{\rm orb}$ decreases. This is a direct consequence of the declining mass transfer rate as systems evolve toward the `period minimum', hence reducing the accumulation of matter in the disc and increasing the time to the next superoutburst. We also showed that the distance between the secondary's photosphere and the L1 point, directly influencing $\dot{M_2}$, may be the factor that distinguish the three SU UMa subclasses from each other. By comparing AR Pic, QW Ser and V521 Peg to other systems in the orbital period ($P_{\rm orb}$) - supercycle ($P_{\rm sc}$) relation, we have determined that they are still evolving toward the `period minimum' and that V521 Peg is close to evolving into a WZ Sge system. 

\section*{Acknowledgements}

Firstly, we thank the anonymous referee for their valuable comments and suggestions. This paper uses observations made at the Boyden Observatory and the SAAO, and from the CRTS, ASAS-3 and ASAS-SN archives, and the AAVSO International Database. We thank all the observers, in particular those who observed the Aug/Sep 2017 superoutburst of V521 Peg: J.P. Marais (University of the Free State), Karl Walke (WLK; AAVSO), Tam\'as Tordai (TRT; Magyar Csillag\'aszati Egyes\"ulet - V\'altoz\'ocsillag Szakcsoport (MCSE-VS) and AAVSO), Roger Pickard (PXR; British Astronomical Association - Variable Star Section (BAA-VSS) and AAVSO) and Stephen Brincat (BSM; AAVSO). We express our appreciation to the National Research Foundation (NRF) for financial assistance. The CRTS is supported by the U.S. National Science Foundation under grant AST-0909182.
 
\section*{Data availability}
The data underlying this article are available in Zenodo, at https://doi.org/10.5281/zenodo.6479204


\bibliographystyle{mnras}
\bibliography{references.bib} 


\bsp	
\label{lastpage}
\end{document}